\documentclass[a4paper,english,reprint,aps,pre,showpacs]{revtex4-1}
\usepackage[T1]{fontenc}
\usepackage[latin9]{inputenc}
\usepackage{array}
\usepackage{multirow}
\usepackage{amsmath}
\usepackage{amssymb}
\usepackage{graphicx}
\usepackage{esint}

\makeatletter

\providecommand{\tabularnewline}{\\}

\@ifundefined{textcolor}{}
{%
 \definecolor{BLACK}{gray}{0}
 \definecolor{WHITE}{gray}{1}
 \definecolor{RED}{rgb}{1,0,0}
 \definecolor{GREEN}{rgb}{0,1,0}
 \definecolor{BLUE}{rgb}{0,0,1}
 \definecolor{CYAN}{cmyk}{1,0,0,0}
 \definecolor{MAGENTA}{cmyk}{0,1,0,0}
 \definecolor{YELLOW}{cmyk}{0,0,1,0}
}

\makeatother

\usepackage{babel}
\begin{document}

\title{Free energies of molecular clusters determined by guided mechanical
disassembly }

\author{Hoi Yu Tang and Ian J. Ford}

\address{Department of Physics and Astronomy and London Centre for Nanotechnology,
University College London, Gower Street, London WC1E 6BT, United Kingdom}
\begin{abstract}
The excess free energy of a molecular cluster is a key quantity in
models of the nucleation of droplets from a metastable vapour phase;
it is often viewed as the free energy arising from the presence of
an interface between the two phases. We show how this quantity can
be extracted from simulations of the mechanical disassembly of a cluster
using guide particles in molecular dynamics. We disassemble clusters
ranging in size from 5 to 27 argon-like Lennard-Jones atoms, thermalised
at 60~K, and obtain excess free energies, by means of the Jarzynski
equality, that are consistent with previous studies. We only simulate
the cluster of interest, in contrast to approaches that require a
series of comparisons to be made between clusters differing in size
by one molecule. We discuss the advantages and disadvantages of the
scheme and how it might be applied to more complex systems.
\end{abstract}


\maketitle

\section{\label{sec:Introduction}Introduction}

The formation of droplets from a metastable vapour phase is a commonplace
event in nature, but so far it has resisted quantitative analysis,
despite repeated attention \cite{kashc00,ford04rev,Vehkamaki-book2006,Kalikmanov-book2013}.
The phenomenon plays a role in atmospheric aerosol and cloud formation
\cite{Zhang12,Kulmala14}, as well as in industrial processes \cite{Agranovski10,turbine1}.
Theoretical analysis often begins with the Becker-Döring equations
\cite{becker35} that describe changes in the populations $n_{i}$
of clusters of $i$ molecules brought about by the processes of gain
and loss of single molecules, or monomers. They take the form
\begin{equation}
dn_{i}/dt=\beta_{i-1}n_{i-1}+\alpha{}_{i+1}n_{i+1}-\left(\beta_{i}+\alpha_{i}\right)n_{i},\label{eq:0}
\end{equation}
where $\beta_{i}$ and $\alpha_{i}$ are growth and evaporation rates,
respectively. The rate of nucleation $J$ of droplets from a metastable
vapour phase may then be expressed as \cite{ford97}
\begin{equation}
J=n_{1}\beta_{i^{*}}Z\exp\left[-\left(\phi(i^{*})-\phi(1)\right)/kT\right],\label{eq:CNT_nucleation_rate}
\end{equation}
where $k$ is the Boltzmann constant, $T$ is the temperature, $i^{*}$
is the size of the critical cluster, defined to have equal probabilities,
per unit time, of molecular gain or loss, and $Z$ is the Zeldovich
factor that accounts for the nonequilibrium nature of the kinetics
\cite{zeldovich42}. We shall refer to $\phi(i)$ as the thermodynamic
work of formation of a cluster of $i$ particles (or $i$-cluster)
starting from the metastable vapour phase. A range of nomenclature
is used for this quantity in the nucleation theory literature: the
work of formation was denoted by $\epsilon(i)$ in \cite{ford97},
and elsewhere the same, or a very similar quantity has been labelled
as $\Delta F$, $\Delta G$ or $\Delta W$, for example.

We note that $\phi$ in the nucleation rate expression has both a
kinetic and a thermodynamic interpretation \cite{ford04}. The quantity
$\phi(i)-\phi(1)$ can be expressed in terms of ratios of cluster
growth and evaporation rates:
\begin{equation}
\phi(i)-\phi(1)=-kT\sum\limits _{j=2}^{i}\ln\frac{\beta_{j-1}}{\alpha_{j}},\label{eq:nucleation_barrier_MC}
\end{equation}
but $\phi$ is also related to the grand potential $\Omega_{s}(i)=F(i)-i\mu_{s}$
of an $i$-cluster at the chemical potential $\mu_{s}$ of the saturated
vapour \cite{ford97}:
\begin{equation}
\phi(i)=\Omega_{s}(i)-ikT\ln S,\label{eq:1a}
\end{equation}
where $F(i)$ is the Helmholtz free energy of the cluster, $S=p_{v}/p_{vs}$
is the vapour supersaturation, and $p_{v}$ and $p_{vs}$ are the
vapour pressure and saturated vapour pressure, respectively. The role
of the grand potential in this context is to specify the equilibrium
population of clusters of size $i$ in a saturated vapour, namely
$n_{i}^{s}=\exp(-\Omega_{s}(i)/kT)$. The nucleation model is completed
by representing the population of monomers as $n_{1}=Sp_{vs}V/kT$,
where $V$ is the system volume, by assuming that the vapour pressure
is dominated by the ideal partial pressure of single molecules.

\begin{figure}
\begin{centering}
\includegraphics[width=1\columnwidth]{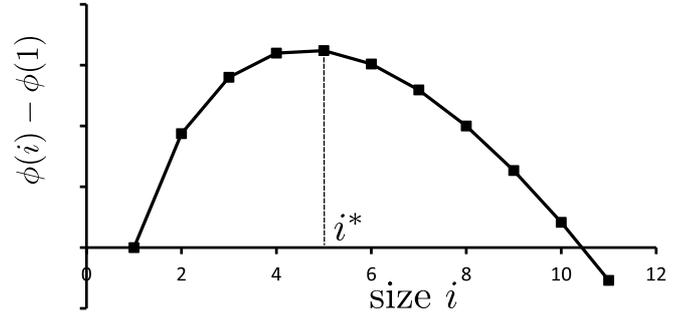}
\par\end{centering}

\protect\caption{\label{fig:work_of_formation}Typical work of formation of a cluster
of $i$ particles with a maximum at the critical cluster size $i^{*}$.}
\end{figure}

In classical nucleation theory (CNT), clusters are viewed as scaled
down versions of macroscopic droplets. According to this approach,
the difference $\phi(i)-\phi(1)$ is replaced by $\phi(i)$ alone
with
\begin{equation}
\phi(i)\approx\phi_{{\rm cl}}(i)=\gamma A(i)-ikT\ln S,\label{eq:1b}
\end{equation}
where $\gamma$ is the surface tension of a planar interface between
vapour and condensate, and $A(i)$ is the surface area of a cluster
represented as a sphere with a density equal to that of the bulk condensed
phase. The work of formation is a combination of a free energy cost
of forming the interface, and a free energy return proportional to
the number of molecules in the cluster (or proportional to its volume
since the condensed phase density is taken to be a constant). The
neglected $\phi(1)$ term might be represented by $\gamma A(1)-kT\ln S$,
which leads to the internally consistent classical theory \cite{girshick90}.

The cluster size dependence of the CNT work of formation is illustrated
in Figure \ref{fig:work_of_formation}. It represents a thermodynamic
barrier, with a maximum at the critical size, that limits the natural
tendency for small molecular clusters to grow into large droplets
when exposed to a supersaturated vapour. CNT has been modified in
several ways, for example by introducing a size-dependent surface
tension \cite{tolman49} or by introducing compatibility with nonideal
vapour properties \cite{laaksonen94,kalikmanov95}.

More fundamentally, the ratio of kinetic coefficients $\beta_{j-1}/\alpha_{j}$
might be evaluated using an underlying microscopic model for all clusters
up to the critical size and beyond \cite{ford04}, and the work of
formation determined through Eq. (\ref{eq:nucleation_barrier_MC}).
It may be shown that
\begin{equation}
\beta_{j-1}/\alpha_{j}=S\exp\left[-[\Omega_{s}(j)-\Omega_{s}(j-1)]/kT\right],\label{eq:thermo-kinetic}
\end{equation}
which shifts attention to the free energy difference $F(j)-F(j-1)$
associated with the addition of a molecule to a $(j-1)$-cluster.
Computing these differences is the basis of an approach has been used
extensively in calculations of cluster free energies \cite{leebarkerabraham1973,haleward1982,tenwoldefrenkel1998,ohzeng1999,kusakaoxtoby2000,chensiepmannklein2001,merikantovehkamakizapadinsky2004}.
But nucleation is actually controlled by the properties of clusters
near the critical size, and one drawback of computing the differences
$F(j)-F(j-1)$ is that the predicted nucleation rate could be susceptible
to the accumulation of errors in evaluating such a sequence.

In this paper, we describe a computational method for directly obtaining
the cluster free energy without the need to perform calculations for
a sequence of smaller clusters. We consider the following representation
of the work of formation of a cluster minus that of a monomer:
\begin{equation}
\phi(i)-\phi(1)=F_{s}(i)-(i-1)kT\ln S.\label{eq:nucleation_barrier_contributions}
\end{equation}
We shall refer to $F_{s}(i)$ as the cluster excess free energy, though
more accurately it is a \emph{difference} between the excess free
energies of an $i$-cluster and a monomer \cite{ford97}. It is `excess'
in that it represents the free energy required to carve a cluster
out of a bulk condensed phase, or equivalently to assemble it out
of saturated vapour. It may be associated with the thermodynamic cost
of creating an interface, which is why in CNT it is modelled by a
surface term, and why we have given it a suffix $s$.

Our approach centres on \emph{disassembling} a cluster into its component
molecules using guided molecular dynamics in order to calculate the
cluster excess free energy directly. The method employs the Jarzynski
equality \cite{jarzynski97,jarzynski97_master,SpinneyFordChapter12}
and we provide details in Section \ref{sec:intro_guided_MD}, including
a comparison with the related method of thermodynamic integration.
Tests of the method where we separate a dimer according to a variety
of protocols are described in Appendix \ref{sec:2ar_test}. The disassembly
of argon-like Lennard-Jones clusters is presented in Section \ref{sec:argon_cluster_disassembly}
and we compare our results with those obtained from Monte Carlo studies
by Barrett and Knight \cite{barrett08} and Merikanto $et\, al.$
\cite{merikanto2006,merikanto2007}. These studies gave consistent
excess free energies, though they were not in agreement with experiments
by Iland $et\, al.$ \cite{strey07}. We conclude with a discussion
of the advantages and disadvantages of the approach compared with
other treatments in Section \ref{sec:conclusions}.

\section{\label{sec:intro_guided_MD}Guided molecular dynamics simulations}

\subsection{Fundamentals of the method}

We study the dynamical evolution of a cluster against a background
of external manipulation. The cluster particles are harmonically tethered
to a set of artificial `guide particles', which lie initially at the
origin but after a period of system equilibration are programmed to
move apart, driving cluster disassembly. The strength of the tether
forces is initially quite weak, in order to disturb the properties
of the cluster as little as possible. Later, the tethers can be strengthened
in order to guide the separation process more firmly, and to prevent
the atoms from interacting with each other once the final guide particle
positions have been reached. The mechanical work of the disassembly
can then be related to the change in Helmholtz free energy.

\begin{figure}
\begin{centering}
\includegraphics[width=1\columnwidth]{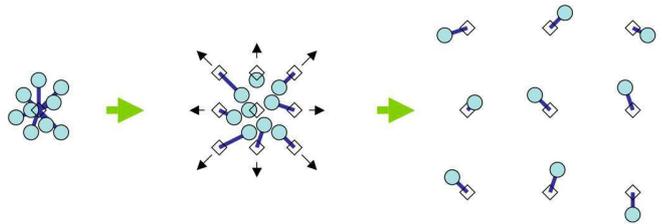}
\par\end{centering}

\protect\caption{\label{fig:clus_sep_diag} Guided disassembly process for an $i$-cluster.
The real particles (circles) are initially weakly tethered to the
guide particles (diamonds). The latter drift apart and the tethers
gradually tighten leading to $i$ independent, tethered particles
upon completion of the process.}
\end{figure}

The masses of the guide particles are taken to be very much greater
than those of the cluster particles. This essentially fixes the trajectories
of the guide particles in the molecular dynamics, in accordance with
the velocities assigned to each at the beginning of the disassembly
process. By choosing guide particle velocities, simulation times and
a time-dependent tethering force, a range of cluster disassembly protocols
can be explored. A simple illustration of the process is shown in
Figure \ref{fig:clus_sep_diag}.

We shall consider clusters of argon-like atoms interacting through
Lennard-Jones potentials, and so we shall refer to the cluster particles
as atoms. We equilibrate this system under the influence of the tethers
for a suitable period, the duration of which will depend upon the
cluster size and the desired temperature. A further molecular dynamics
simulation is performed and from this trajectory we select initial
configurations for cluster disassembly. In order that the configurations
should represent a bound structure, we employ a Stillinger cluster
condition \cite{stillinger63} in the selection, allowing a separation
of no more than $1.5\,\sigma_{{\rm ArAr}}$ between an atom and its
nearest neighbour, where $\sigma_{{\rm ArAr}}$ is the usual Lennard-Jones
range parameter. Such a Stillinger condition has been used in previous
Monte Carlo approaches. The cluster definition is an important ingredient
of a modelling strategy \cite{ford04rev}, and deserves careful consideration,
but here we shall use this simple criterion for convenience.

The simulations were performed using the DL\_POLY \cite{dlpoly} molecular
dynamics package, with modifications to the source code to implement
the time-dependent harmonic tether potentials. We include a physical
heat bath of helium-like Lennard-Jones atoms thermalised using a Nosé-Hoover
thermostat \cite{tangford2006}. We could instead have implemented
a thermostat that acts on the cluster itself, but chose not to in
order to achieve as natural a thermalisation as possible during the
nonequilibrium processing.

\subsection{Work performed on a system}

Given an external control parameter $\lambda$ in a Hamiltonian $H(\lambda)$,
the work $W$ done on a system due to the evolution of $\lambda$
over a finite time period may be written
\begin{equation}
W=\int\frac{d\lambda}{dt}\frac{\partial H(\lambda)}{\partial\lambda}dt.\label{eq:work_1steqn}
\end{equation}
For example, consider the Hamiltonian $H_{1}$ of a single guided
atom of mass $m$:
\begin{equation}
H_{1}=\frac{\boldsymbol{p}^{2}}{2m}+\frac{1}{2}\kappa(t)\left[\boldsymbol{x}(t)-\mathbf{X}(t)\right]^{2},\label{eq:oscill_hamilt}
\end{equation}
where $\boldsymbol{p}$ is the momentum, $\kappa(t)$ is the time-dependent
tethering force or spring constant, $\boldsymbol{x}(t)$ is the atomic
position and $\mathbf{X}(t)$ is the guide position. For a set of
guided atoms, each controlled by a Hamiltonian $H$ containing terms
of the form given in Eq. (\ref{eq:oscill_hamilt}) supplemented by
interparticle interactions, $\kappa(t)$ and $\mathbf{X}(t)$ play
the role of $\lambda$ and the work $W$ performed on the set is
\begin{equation}
\begin{aligned} & \int\limits _{0}^{\tau}\frac{d\kappa(t)}{dt}\frac{\partial H(\kappa,\{\mathbf{X}_{k}\})}{\partial\kappa}dt+\sum_{j}\int\limits _{0}^{\tau}\frac{d\mathbf{X}_{j}(t)}{dt}\frac{\partial H(\kappa,\{\mathbf{X}_{k}\})}{\partial\mathbf{X}_{j}}dt\\
 & =\frac{1}{2}\int\limits _{0}^{\tau}\frac{d\kappa(t)}{dt}\sum_{j=1}^{i}\left[\boldsymbol{x}_{j}(t)-\mathbf{X}_{j}(t)\right]^{2}dt\\
 & -\int\limits _{0}^{\tau}\kappa(t)\sum_{j=1}^{i}\left[\boldsymbol{x}_{j}(t)-\mathbf{X}_{j}(t)\right]\cdot\mathbf{V}_{j}(t)dt,
\end{aligned}
\label{eq:thermo_integr}
\end{equation}
where $\tau$ is the length of the molecular dynamics simulation,
and $\mathbf{V}_{j}(t)$ is the velocity of the guide particle associated
with the $j^{{\rm th}}$ atom, defined as $\mathbf{V}_{j}(t)=d\mathbf{X}_{j}(t)/dt$.
The first term in Eq. (\ref{eq:thermo_integr}) arises from the time
dependence of the spring constant, and the second term is simply the
conventional force times distance expression. It should be noted that
all tethers within the system are characterised by the same spring
constant, although more elaborate protocols could be imagined.

\subsection{\label{sub:Jarzynski}The Jarzynski equality}

If we were able to perform an extremely slow, quasistatic process,
then the mechanical work done would be equal to the difference in
Helmholtz free energy between the initial and final equilibrium states.
However, quasistatic processes are unfeasible in finite time molecular
dynamics simulations and according to the second law \cite{fordbook},
the average of the work done (as a result of a time-dependent change
in the Hamiltonian of the system), performed over many realisations
of a nonquasistatic process (indicated by angled brackets), will always
be an overestimate of the free energy change, $\left\langle W\right\rangle >\Delta F$,
allowing us only to infer an upper limit to $\Delta F$.

However, the Jarzynski equality \cite{jarzynski97,jarzynski97_master}
\begin{equation}
\left<\exp\left(-W/kT\right)\right>=\exp\left(-\Delta F/kT\right)\label{eq:Jarzynski_Equality}
\end{equation}
allows us to do better. For this identity to hold, the system must
begin in thermal equilibrium, but need not remain so as the Hamiltonian
changes during the simulation. Exploiting the work done in a nonequilibrium
process is a powerful strategy for calculating cluster surface free
energies and numerous computational studies \cite{hendrix2001,hummer2002,hu2002,rodriguezgomez2004,lua2005,dhar2005,palmieri2007}
as well as experiments \cite{hummer2001,ritort2002,liphardt02,collin2005,douarche2005,joubaud2007}
have achieved this with the help of the Jarzynski equality. Systems
studied include argon-like Lennard-Jones fluids, ion-charging in water,
ideal gases confined to a piston, and one-dimensional polymer chains.
Nevertheless, there are distinct aspects of this strategy for analysing
the controlled disassembly of a cluster that need to be explored.

The Jarzynski equality ought to recover the free energy difference
regardless of the nature of the evolution between initial and final
Hamiltonians, but computed results might still depend upon the rate
of the process as a consequence of a limited sampling of system trajectories
in finite simulations \cite{hummer2001}. We might expect `slow' processes
that gently pull a cluster apart to generate a narrower distribution
of work compared with `fast' processes that are violent and highly
dissipative. A balance must therefore be struck between the poorer
convergence of fast simulations and the demand for computational resources
required for slow simulations.

Furthermore, a consequence of the exponential averaging in the Jarzynski
equality is that occasional values of work that are well below the
average, arising from unusual trajectories, can sometimes distort
the extracted free energy change. This is a consequence of insufficient
sampling of the system trajectories and so we need to give careful
attention to the statistical errors.

We have explored the outcomes of various guiding protocols, and the
robustness of the Jarzynski equality in the face of limited statistics,
in a test case of the separation of a dimer, for which the free energy
change is easily calculable. These studies are described in Appendix
\ref{sec:2ar_test}. We have used similar protocols to study the disassembly
of larger clusters, which is described in Section \ref{sec:argon_cluster_disassembly}.

\subsection{Comparison with thermodynamic integration}

The method bears some similarity to thermodynamic integration, where
the strength of the interparticle interactions is evolved over a sequence
of equilibrium calculations in order to compare the system in question
with another that has a known free energy \cite{Kirkwood35,straatsma86,miller00,schilling09}.
The basic relationship $\Delta F=\int\langle\partial H(\lambda)/\partial\lambda\rangle d\lambda$
is analogous to Eq. \eqref{eq:work_1steqn}. The reference system
for clusters might, for example, be a set of noninteracting particles
held together through the retention of the constraining cluster definition.
Or indeed the cluster definition could be changed progressively along
with the interactions in order to reach a more convenient final state,
perhaps noninteracting particles inside a sphere.

However, there are some important differences. In our approach it
is the tether potentials that change with time, not the interparticle
interactions, and our reference system is a set of independent harmonic
oscillators, not an ideal gas. Furthermore, we conduct the evolution
by nonequilibrium molecular dynamics rather than by moving through
a sequence of equilibrium ensembles, and we only need to impose a
cluster definition when selecting the initial configurations, not
throughout the evolution. An abrupt removal of the cluster definition
constraint is acceptable in a nonequilibrium evolution, when the results
are processed using the Jarzynski equation, but it would not be appropriate
during a sequence of equilibrium calculations.

\section{\label{sec:argon_cluster_disassembly}Argon cluster disassembly}

\subsection{Preliminaries}

We have investigated the disassembly of clusters consisting of 5,
10, 15, 20 and 27 argon-like atoms in order to obtain their excess
free energies. Scaling up the guided molecular dynamics simulations
from the test case of dimer separation is fairly straightforward.
We perform simulations in a cubic cell with edge lengths of 100~Å,
so that the initial clusters and the final disassembled configurations
may be easily accommodated. We employ Lennard-Jones interaction potentials
for each species (see Table \ref{tab:LJ-parameters1}) and the helium
temperature is set at 60 K in order to facilitate a comparison with
the Monte Carlo studies by Barrett and Knight \cite{barrett08} and
Merikanto $et\, al.$ \cite{merikanto2006,merikanto2007}, as well
as the experimental studies of Iland $et\, al.$ \cite{strey07}.

However, converting the free energy change associated with disassembly
into an excess free energy requires some careful consideration of
the statistical mechanics of tethered and free molecular clusters.
We require the excess free energy of a cluster that is free to move
anywhere inside a system volume, but our initial state is a cluster
tethered to guide particles at the origin. The free energy change
that emerges from our calculations will correspond to the disassembly
of a cluster whose centre of mass explores a region around the origin,
and furthermore, one that possesses energy due to the tethers in addition
to that of the physical interactions between the atoms. These matters
are discussed in detail in Appendix B.

\begin{figure}
\begin{centering}
\includegraphics[width=1\columnwidth]{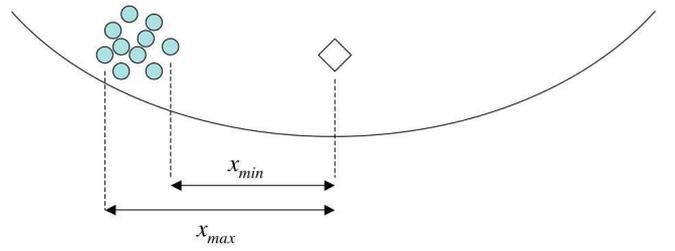}
\par\end{centering}

\protect\caption{\label{fig:kappai_criterion} The difference in tether energy across
a cluster configuration is given in terms of the maximum and minimum
separations between an atom and its guide particle. The circles depict
the argon atoms, while the diamond represents the position of all
of the guides at the origin of the cell.}
\end{figure}

The energetic perturbation of the cluster by the tethers can be reduced
by choosing a small force constant. We take the view that the mean
variation in tethering energy of an atom, as it explores different
regions of the cluster during the equilibrated trajectory, should
not exceed the thermal energy $kT$, or
\begin{eqnarray}
\frac{1}{2}\kappa_{i}\langle x_{{\rm max}}^{2}-x_{{\rm min}}^{2}\rangle & < & kT,\label{eq:kappai_criterion}
\end{eqnarray}
where $x_{{\rm max}}$ and $x_{{\rm min}}$ are, respectively, the
maximum and minimum separations between an atom and its guide particle
in a configuration (see Figure \ref{fig:kappai_criterion}). This
criterion may also be expressed as $\xi=\kappa_{i}\langle x_{{\rm max}}^{2}-x_{{\rm min}}^{2}\rangle/(2kT)<1$.

From the equilibrated molecular dynamics trajectory, we select, for
disassembly, a set of `valid' cluster configurations that satisfy
the Stillinger cluster definition \cite{stillinger63}, but this can
be quite difficult for the smaller clusters at 60 K. Tethering the
atoms keeps them closer together and more likely to form valid configurations.
We therefore choose a tethering strength that satisfies the condition
on $\xi$, but also helps to produce sufficient valid cluster configurations.
The initial value of the tethering force constant was taken to be
$\kappa_{i}=0.01\,\mbox{kJ mol}^{-1}\mbox{Å}^{-2}$, which gives $\xi\sim0.6-0.9$
for the five sizes of argon cluster studied. Table \ref{tab:i-Ar_equil_data}
shows the duration of the equilibrated cluster trajectory, the number
of valid cluster configurations identified from candidates selected
at intervals of 100~ps from the equilibrated molecular dynamics trajectory,
and the ratio $\xi$ characterising the suitability of the tethering
force constant.

\begin{table}
\centering{}%
\begin{tabular}{cccc}
\hline
$i$ & Duration/ns & Valid configurations & $\xi$\tabularnewline
\hline
\hline
5 & 1000 & 152 & 0.604\tabularnewline
10 & 250 & 411 & 0.745\tabularnewline
15 & 225 & 1070 & 0.799\tabularnewline
20 & 150 & 905 & 0.847\tabularnewline
27 & 150 & 1020 & 0.922\tabularnewline
\hline
\end{tabular}\protect\caption{\label{tab:i-Ar_equil_data}The duration of equilibrated cluster trajectories
at 60~K, as well as the number of valid cluster configurations selected
at each size. The ratio $\xi$ characterises the perturbation to the
cluster energy due to the tether potentials.}
\end{table}

Having obtained initial cluster configurations for the five sizes
of cluster, the next stage is to disassemble them by a combination
of guide particle motion and tether tightening. A range of separation
times $t_{{\rm sep}}$ is explored, with the larger and more stable
clusters expected to require longer disassembly processes in order
to provide accurate estimates of the free energy change. As in the
dimer calculations described in Appendix A, we use a tethering strength
that strengthens in time according to Eqs. (\ref{eq:spring_mod-1}),
with a final value of $\kappa_{f}=0.05\,\mbox{kJ mol}^{-1}\mbox{Å}^{-2}$.

The terminal positions for the guide particles are chosen from a $3\times3\times3$
grid with spacing of $33.33\,\mbox{Å}$. The largest cluster considered
contains 27 argon atoms so after the process of disassembly, the tethered
atoms move around each point on this grid. For smaller systems, the
same grid of final guide positions is adopted, but employing only
as many points as are necessary for the cluster in question. With
initial guide positions at the origin and final positions defined
in this way, it is straightforward to calculate the necessary drift
velocities of the guide particles for a given separation time. Applying
the Jarzynski procedure to the distribution of performed work then
gives us the estimated free energy change $\Delta F$ associated with
the disassembly of a cluster.

However, as mentioned previously, this free energy difference will
only correspond to the disassembly of a tethered $i$-cluster, rather
than of a freely translating, undistorted cluster. Furthermore, by
necessity we obtain free energies of systems of \emph{distinguishable}
atoms in molecular dynamics, and we need to make an indistinguishability
correction. An analysis of the thermodynamics is required in order
to extract the excess free energy of an $i$-cluster from the free
energy of disassembly, and the details are given in Appendix B. It
turns out that we can write $F_{s}(i)=\sum_{k=1}^{5}f_{s}^{k}(i)$
with
\begin{eqnarray}
f_{s}^{1}(i) & = & -\Delta F\label{eq:fren30_breakdown1}\\
f_{s}^{2}(i) & = & -ikT\ln\left(\rho_{vs}v_{{\rm HO}}\right)\label{eq:fren30_breakdown2}\\
f_{s}^{3}(i) & = & kT\ln\left(\rho_{vs}v_{c}\right)\label{eq:fren30_breakdown3}\\
f_{s}^{4}(i) & = & -\frac{3i\kappa_{i}}{10}\left(\frac{3iv_{l}}{4\pi}\right)^{2/3}\label{eq:fren30_breakdown4}\\
f_{s}^{5}(i) & = & kT\ln i!\label{eq:fren30_breakdown5}
\end{eqnarray}
In the first term the free energy of disassembly $\Delta F$ appears
with a negative sign because it refers to the process of taking a
cluster apart while $F_{s}$ is the free energy of interface formation.
The $f_{s}^{2}$ term arises from relating the final state in the
disassembly process, namely the separated harmonically bound particles,
to the appropriate reference state of a saturated vapour. It represents
the difference in free energy between the tethered particles, each
effectively confined to a volume $v_{{\rm HO}}=\left(2\pi kT/\kappa_{f}\right)^{3/2}$,
and particles in the saturated vapour phase with density $\rho_{vs}$
and volume per particle $1/\rho_{vs}$. The $f_{s}^{3}$ term is the
entropy penalty associated with the initial tethering: the centre
of mass of the cluster is effectively confined to a volume $v_{c}=\left(i\kappa_{i}/(2\pi kT)\right)^{-3/2}$
and needs to be referred to a situation where it is allowed, like
a particle in saturated vapour, to explore a volume $1/\rho_{vs}$.
The $f_{s}^{4}$ term is an approximate expression for the perturbation
in the cluster energy due to the initial presence of the tethers,
where $v_{l}=1/\rho_{l}$ is the volume per particle in the condensed
phase. Finally, $f_{s}^{5}$ converts calculations derived from molecular
dynamics with distinguishable particles into results relevant to a
system of indistinguishable particles.

\subsection{\label{sec:discussion}Results and discussion}

\begin{figure}
\begin{centering}
\includegraphics[width=1\columnwidth]{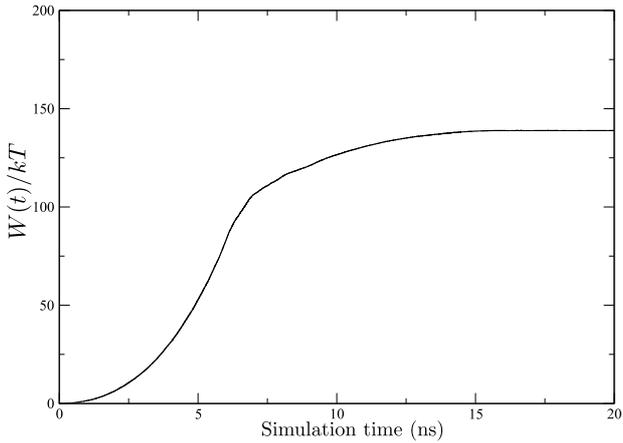}
\par\end{centering}

\protect\caption{\label{fig:27Ar_singlerun_work}A typical history of the work performed
for one realisation of the disassembly of a 27-atom argon cluster
with a separation time of $20\,\mbox{ns}$.}
\end{figure}

A typical example of the work $W(t)$ performed over a disassembly
trajectory of duration 20 ns for a 27-atom cluster is shown in Figure
\ref{fig:27Ar_singlerun_work}. The gradual rise in the work performed
prior to about $5\mbox{ ns}$ represents an accumulation of tethering
energy as the guide particles move away from their initial positions
at the origin. After this time, atoms begin to leave the cluster,
and less work is needed to move the corresponding guides. After about
7 ns, the work rate reduces significantly as the cluster disintegrates
and the guide particles move towards their final positions.

\begin{figure}
\noindent \begin{centering}
\includegraphics[width=0.47\columnwidth]{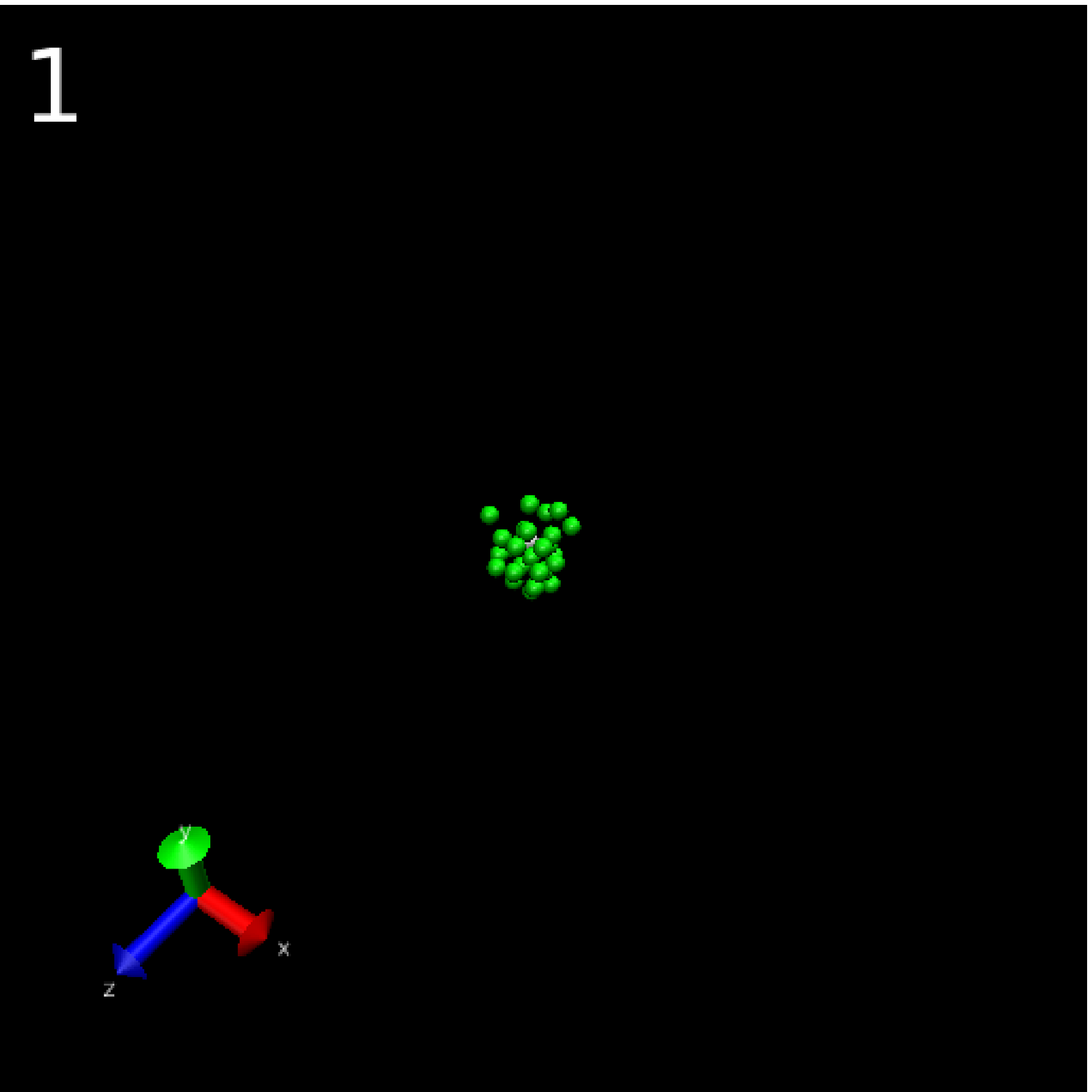} \includegraphics[width=0.47\columnwidth]{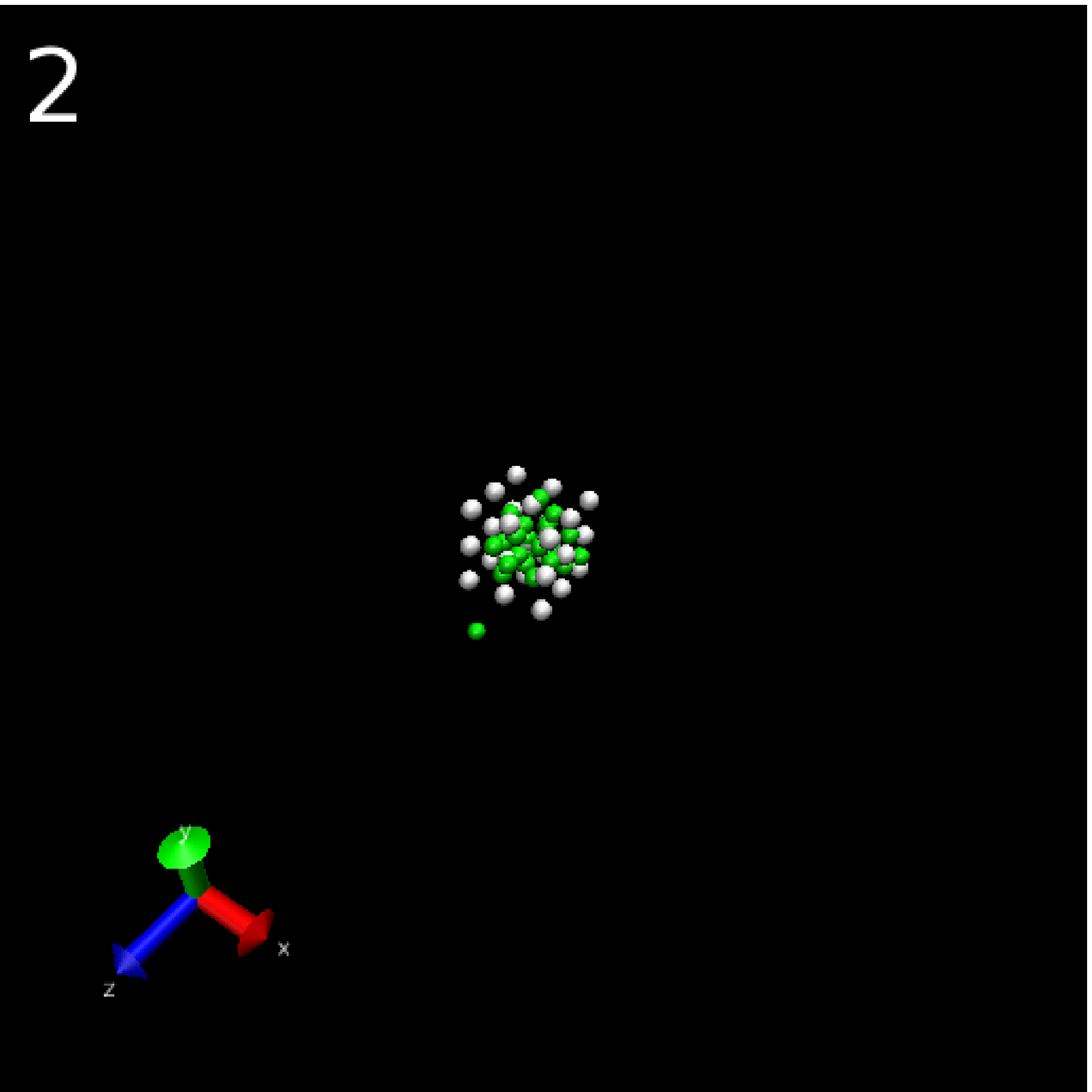}
\par\end{centering}

\noindent \begin{centering}
\includegraphics[width=0.47\columnwidth]{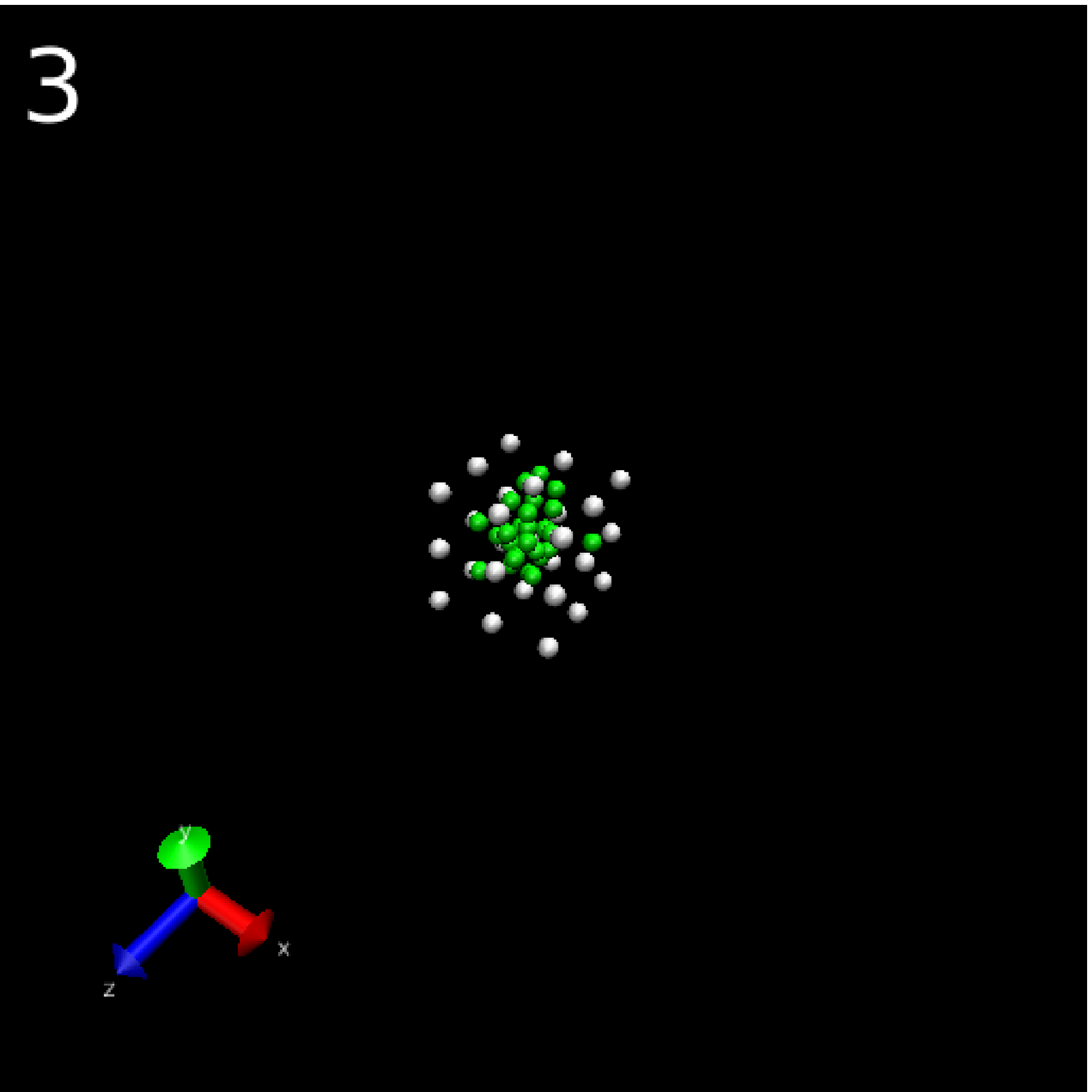} \includegraphics[width=0.47\columnwidth]{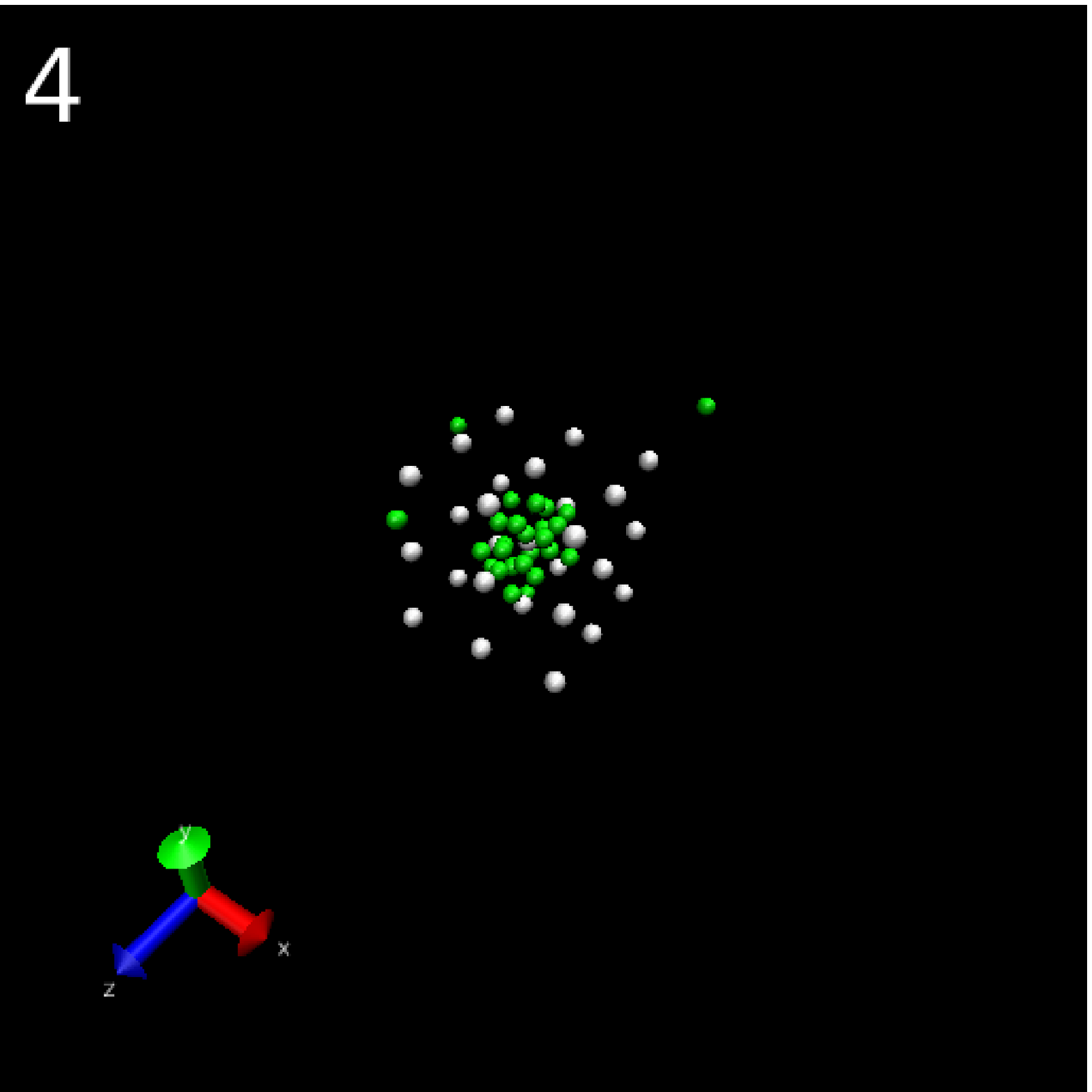}
\par\end{centering}

\noindent \begin{centering}
\includegraphics[width=0.47\columnwidth]{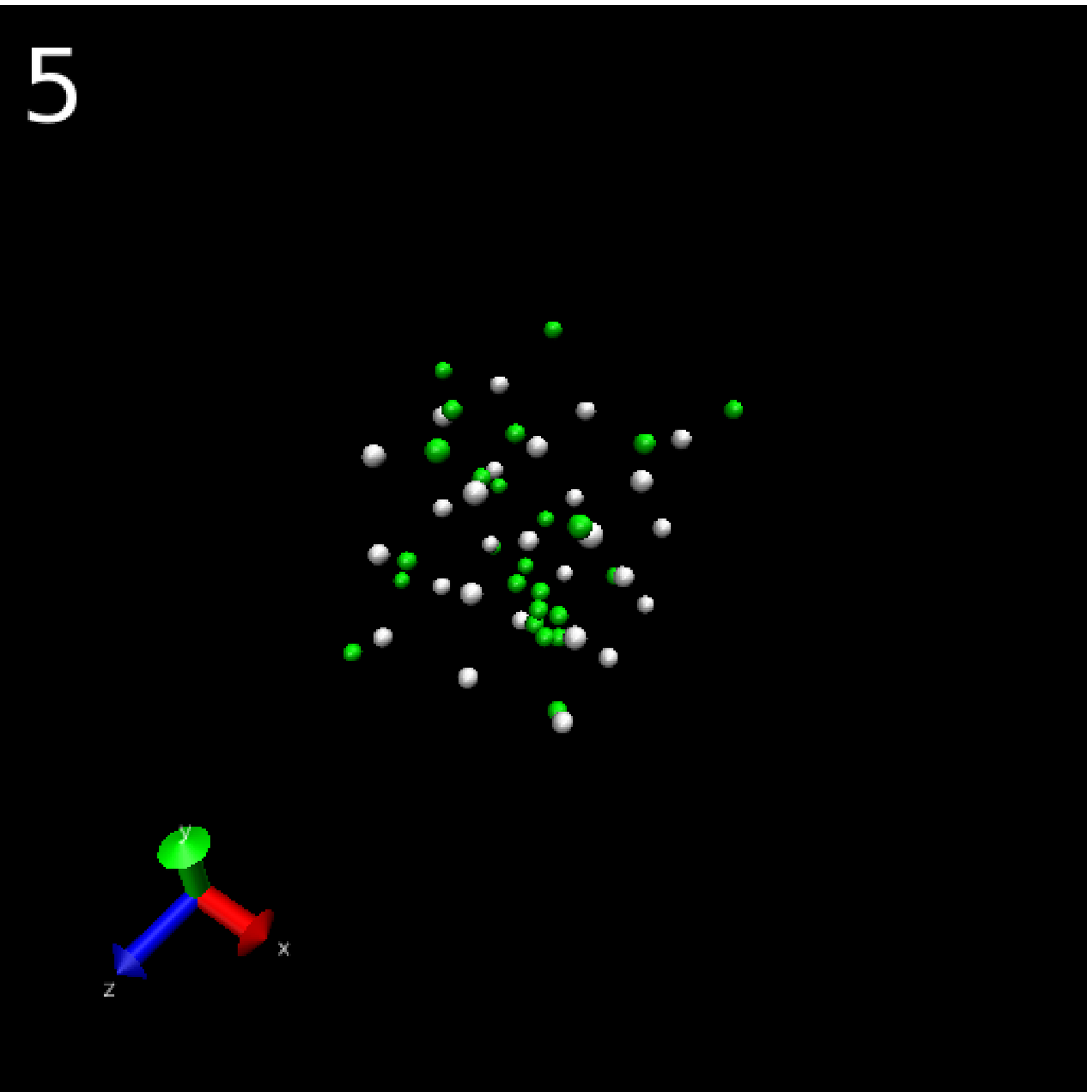} \includegraphics[width=0.47\columnwidth]{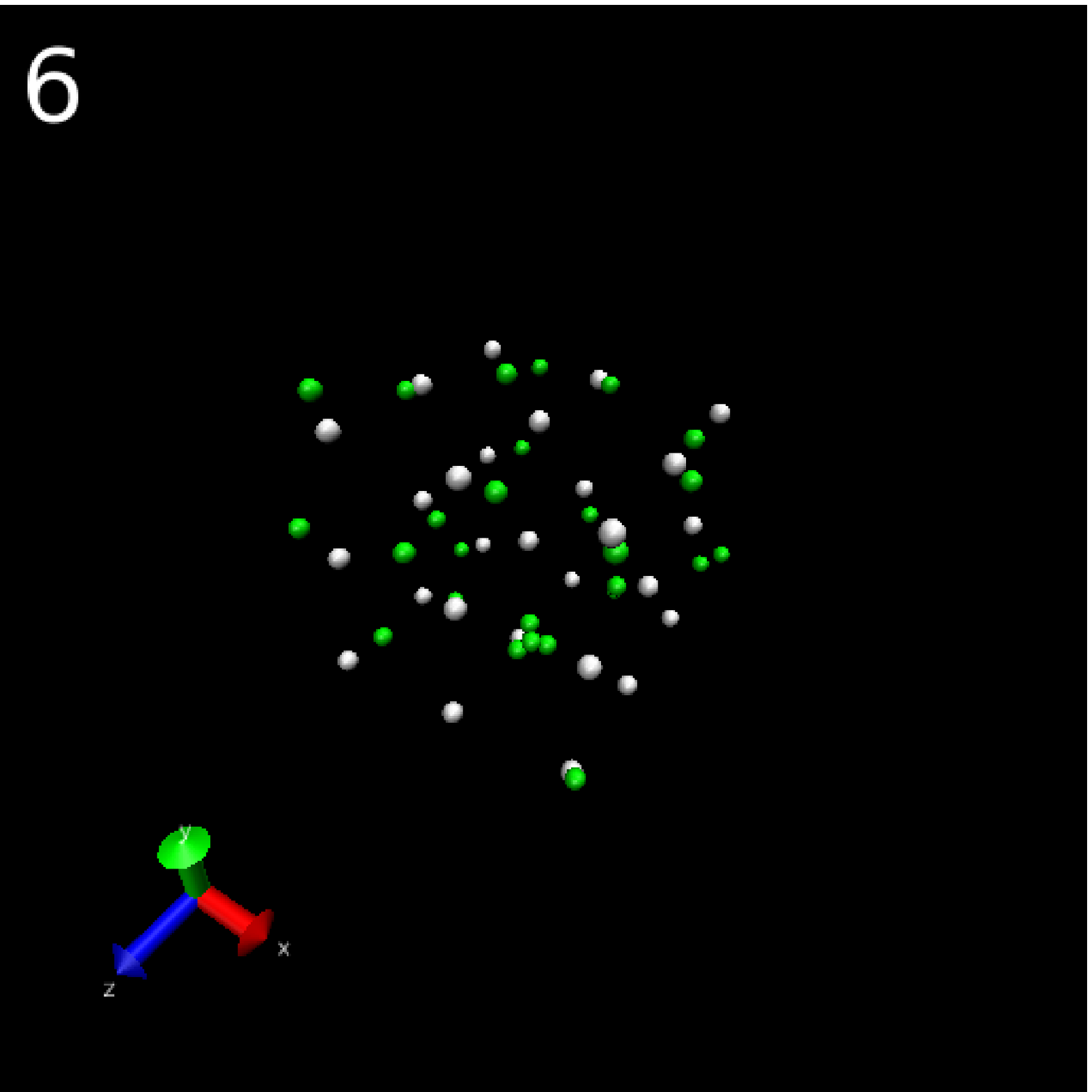}
\par\end{centering}

\noindent \begin{centering}
\includegraphics[width=0.47\columnwidth]{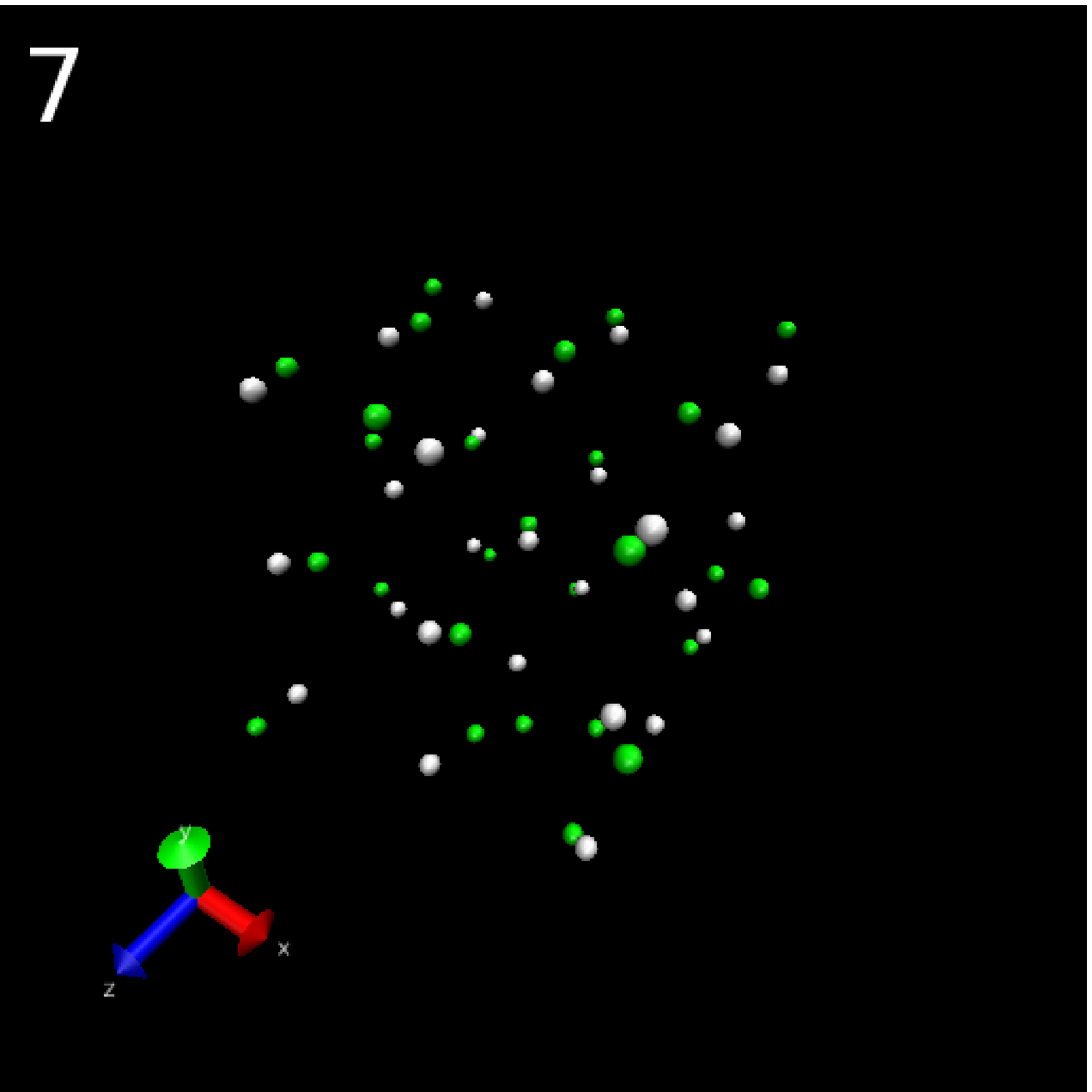} \includegraphics[width=0.47\columnwidth]{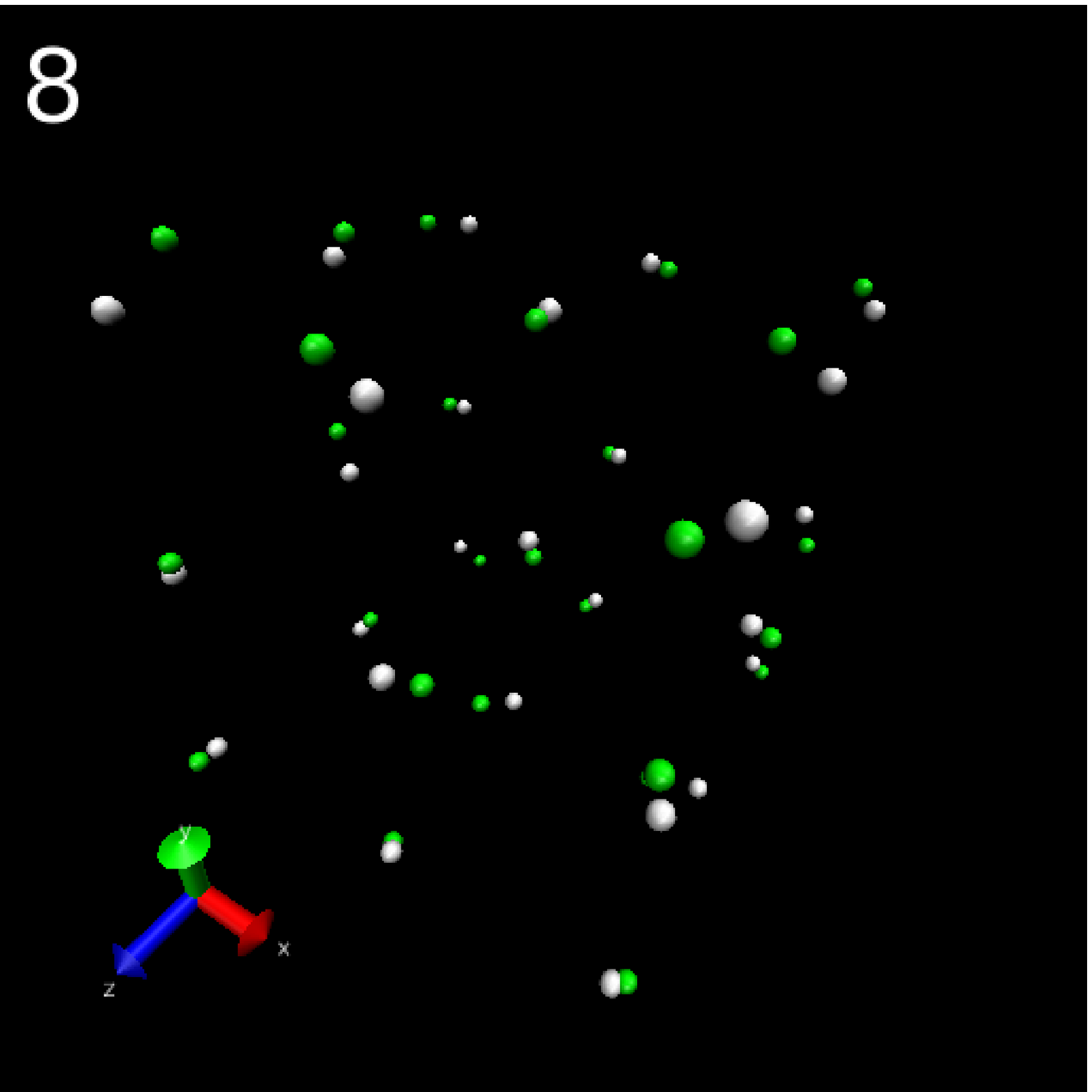}
\par\end{centering}

\protect\caption{\label{fig:27ar_sep_snapshots}Illustration of the disassembly of
a 27-atom argon cluster, with green spheres representing the argon
atoms and lighter spheres the guide particles (helium atoms are not
shown). In frame 1, all the guides lie at the origin of the cell.
By frame 2, the guides have drifted far enough apart for a single
argon atom to escape temporarily from the cluster before rejoining
it in frame 3. In frame 4, several atoms have escaped, but remain
in close proximity to the reduced cluster. A threshold is reached
in frame 5, where many argon atoms break free to leave a fragment
of about five atoms that also soon disintegrates as shown in frame
6. Shortly after, all of the atoms fall into motion about their partner
guide particles which continue along steady paths away from one another
(frames 7 and 8). The reader is encouraged to view movies of the disassembly
provided in the Supplemental Material \cite{suppl}.}
\end{figure}

Visual representations of the disassembly process (see Figure \ref{fig:27ar_sep_snapshots})
provide further insight into the manner in which the clusters are
pulled apart. The onset of cluster disassembly is signalled by the
loss of one or two atoms from the cluster, perhaps only temporarily.
The cluster soon after breaks into several smaller clusters, which
eventually disintegrate into fragments or single atoms. It is rare
to see a complete and sudden disintegration of a cluster, where all
the constituent atoms disassemble together within a short space of
time.

\begin{figure}
\begin{centering}
\includegraphics[width=1\columnwidth]{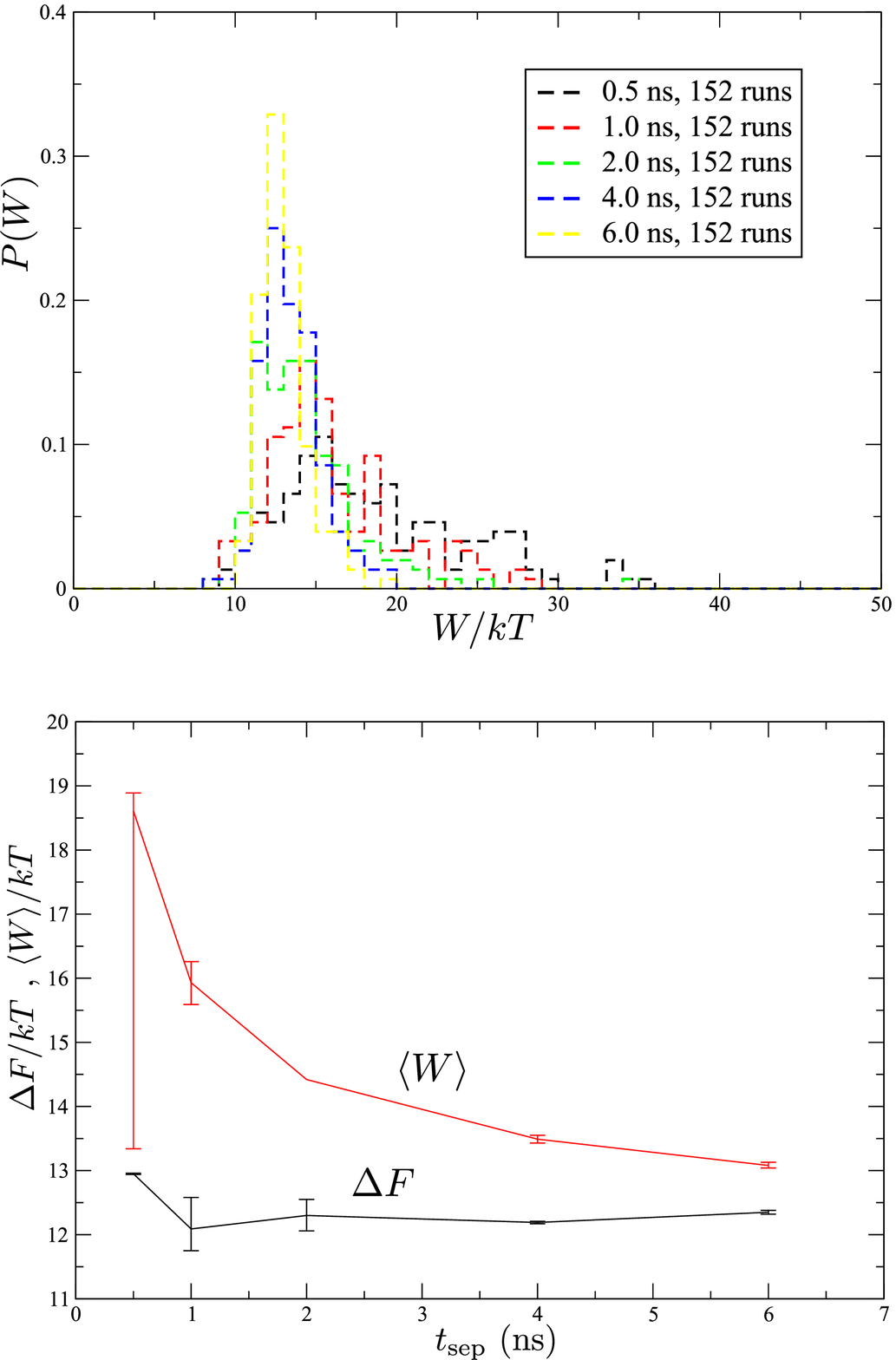}
\par\end{centering}

\protect\caption{\label{fig:5Ar_work.dist_DF} The distribution of work $W$ (top)
for sets of disassembly trajectories for the 5-atom argon cluster,
for a range of separation times. The lower plot shows the mean of
the work $\langle W\rangle$ and the corresponding free energy differences
$\Delta F$ calculated via the Jarzynski equality for each $t_{{\rm sep}}$.}
\end{figure}

\begin{figure}
\begin{centering}
\includegraphics[width=1\columnwidth]{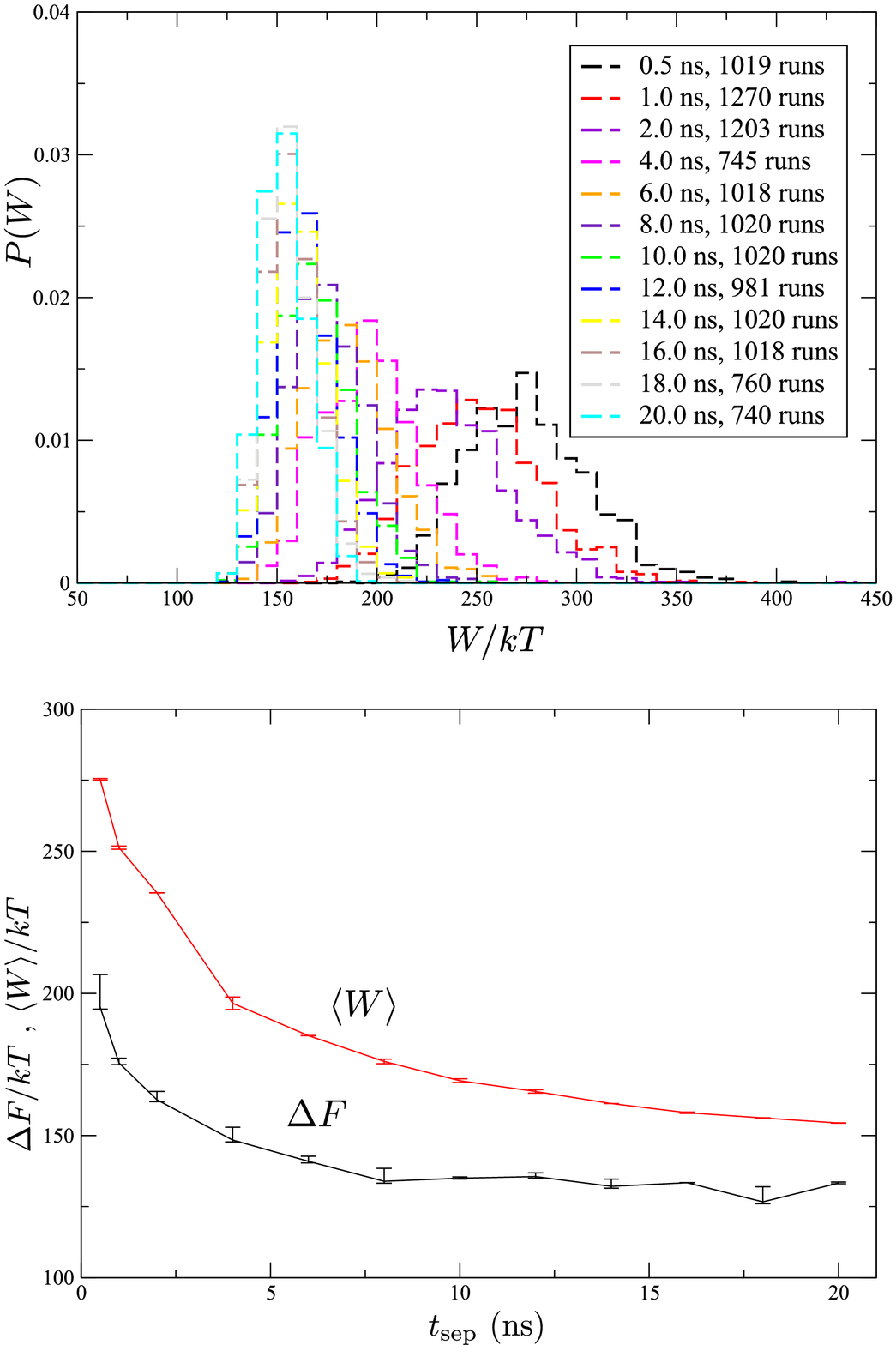}
\par\end{centering}

\begin{centering}

\par\end{centering}

\protect\caption{\label{fig:27Ar_work.dist_DF} Plots similar to those shown in Figure
\ref{fig:5Ar_work.dist_DF} but for the 27-atom argon cluster. }
\end{figure}

Figures \ref{fig:5Ar_work.dist_DF} and \ref{fig:27Ar_work.dist_DF}
show distributions of the work performed in disassembling the 5-cluster
and the 27-cluster, along with estimates of the free energy change,
for separation times between 0.5~ns and 20~ns. As expected, the
work distributions are broader for the processes that are most rapid
(smallest $t_{{\rm sep}}^{-1}$) and hence least quasistatic in nature.
Conversely, the work distributions become narrower, and lead to free
energy changes that presumably provide the most accurate estimates
of the true free energy change, as the rate of separation is reduced.

The free energy change $\Delta F$ for the disassembly of each size
of cluster at the slowest rate studied is shown in Table \ref{tab:excess free energies},
along with the other contributions to the excess free energy $F_{s}$.
We refer to a molecular dynamics study by Baidakov $et\, al.$ \cite{baidakov2007}
to provide values of the saturated vapour density $\rho_{vs}$ and
liquid density $\rho_{l}=1/v_{l}$ of the argon-like Lennard-Jones
fluid at a temperature of 60.31~K.

\begin{table}
\begin{centering}
\begin{tabular}{ccccccccc}
\hline
$i$ & $t_{{\rm sep}}(\mathrm{ns)}$ & $\langle W\rangle$ & $\Delta F$ & $f_{s}^{2}(i)$ & $f_{s}^{3}(i)$ & \multirow{1}{*}{$f_{s}^{4}(i)$} & $f_{s}^{5}(i)$ & $F_{s}(i)$\tabularnewline
\hline
\hline
5 & 6 & 13.08 & 12.35 & 38.94 & -7.79 & -0.41 & 4.79 & 23.18\tabularnewline
10 & 8 & 34.06 & 30.80 & 77.88 & -8.83 & -1.32 & 15.10 & 52.03\tabularnewline
15 & 12 & 62.75 & 53.87 & 116.81 & -9.44 & -2.59 & 27.90 & 78.82\tabularnewline
20 & 16 & 97.37 & 84.07 & 155.75 & -9.87 & -4.18 & 42.33 & 99.97\tabularnewline
27 & 20 & 154.41 & 133.28 & 210.27 & -10.32 & -6.90 & 64.56 & 124.33\tabularnewline
\hline
\end{tabular}
\par\end{centering}

\protect\caption{Results from the slowest set of disassembly simulations for each cluster
size: the mean work $\langle W\rangle$, the free energy of disassembly
$\Delta F$ and the other contributions to the excess free energy
$F_{s}(i)$, all in units of $kT$. \label{tab:excess free energies}}
\end{table}

\begin{figure}
\begin{centering}
\includegraphics[width=1\columnwidth]{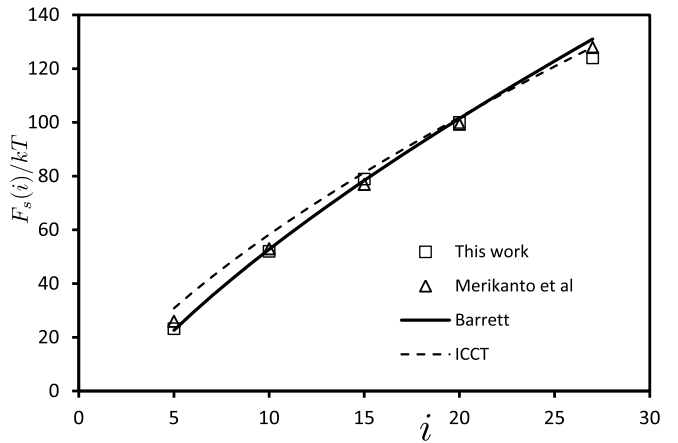}
\par\end{centering}

\protect\caption{Excess free energies for argon-like Lennard-Jones clusters obtained
from disassembly at 60 K are shown as squares and compared with values
obtained in Monte Carlo studies by Barrett and Knight \cite{barrett08}
at $59.88\,\mbox{K}$ (solid line) and Merikanto $et\, al.$ \cite{merikanto2006,merikanto2007}
at $60.18\,\mbox{K}$ (triangles). Also shown is the prediction from
internally consistent classical nucleation theory for a temperature
of $60.31\,\mbox{K}$ (dashed line). \label{fig:compare Ws}}
\end{figure}

Figure \ref{fig:compare Ws} shows our excess free energies $F_{s}(i)$
as a function of cluster size $i$. Statistical errors propagated
from uncertainties in the free energy change $\Delta F$ are similar
to the size of the symbols. We also include corresponding results
from the Monte Carlo studies by Barrett and Knight \cite{barrett08}
and Merikanto $et\, al.$ \cite{merikanto2006,merikanto2007}. Barrett
and Knight employed a Lee-Barker-Abraham cluster definition \cite{leebarkerabraham1973}
while Merikanto $et\, al.$ adopted a Stillinger cluster criterion
similar to ours. The Barrett and Knight calculations are represented
here by $F_{s}^{{\rm BK}}(i)/kT=-\ln q_{i}-(i-1)\ln(\rho_{vs}\sigma_{{\rm ArAr}}^{3})$
with their fitting function $\ln q_{i}=10.5+9.91(i-1)-16.36(i^{2/3}-1)$,
and the Merikanto $et\, al.$ values are derived from their Figure
1 in \cite{merikanto2006}, which we interpret as a plot of $F_{s}^{{\rm M}}(i)/kT-(i-1)\ln S$
with $S=20$. The results of these earlier studies are consistent
with one another, as well as with the excess free energy suggested
by the internally consistent classical theory (ICCT) $F_{s}^{{\rm ICCT}}(i)=\gamma\left(36\pi v_{l}^{2}\right)^{1/3}\left(i^{2/3}-1\right)$,
constructed such that $F_{s}^{{\rm ICCT}}(1)=0$, where $\gamma$
is the surface tension of the planar liquid-vapour interface, again
taken from Baidakov $et\, al.$ \cite{baidakov2007}. It is clear
from Figure \ref{fig:compare Ws} that the calculations presented
in this study are consistent with the previous Monte Carlo results.
This is satisfactory support for the disassembly approach that we
have developed. We note that all three are reasonably well represented
by the ICCT model, which is somewhat surprising.

Note that the construction of a traditional plot of the nucleation
barrier such as Figure \ref{fig:work_of_formation} would require
us to subtract a term $ikT\ln S$ from the excess free energies in
Figure \ref{fig:compare Ws}. Inserting a supersaturation of $30$
would then yield a critical size of about 20, for example.

\section{\label{sec:conclusions}Conclusions}

We have developed a method of guided cluster disassembly in molecular
dynamics, capable of extracting the excess free energy associated
with the formation of a molecular cluster from the saturated vapour
phase. This property is often regarded as a surface term and it plays
a central role in kinetic and thermodynamic models of the process
of droplet nucleation.

After exploring some aspects of the method by separating a dimer,
the technique was applied to the controlled disassembly of Lennard-Jones
argon clusters between 5 and 27 atoms in size. The extracted free
energy of disassembly has been related to the excess free energy of
the cluster through an analysis of the statistical mechanics of free
and tethered clusters. Our calculations for clusters of various sizes
are consistent with previous studies by Barrett and Knight \cite{barrett08}
and Merikanto \emph{et al}. \cite{merikanto2006,merikanto2007}, both
of which require the evaluation of a sequence of free energy differences
between monomer and dimer, dimer and trimer, etc. A Lennard-Jones
microscopic model of argon, within the standard kinetic and thermodynamic
framework of nucleation theory, cannot account for the experimental
argon nucleation data of Iland \emph{et al.} \cite{strey07}, but
we do not speculate here about this disparity.

The approach should be contrasted with methods of free energy estimation
based on thermodynamic integration. In those methods, the strength
of the interparticle interactions is evolved over a sequence of equilibrium
calculations. Our approach also involves the evolution of a Hamiltonian,
but it is the tether potentials that change with time, not the interparticle
interactions. Furthermore, we evolve by nonequilibrium molecular dynamics
rather than studying a sequence of equilibrium ensembles, and we are
only required to apply a cluster definition when selecting the initial
configurations, not during the evolution.

We believe that our process of mechanical disassembly offers an intuitive
understanding of the meaning of the work of formation that plays such
a central role in nucleation theory. We suggest that a direct evaluation
of this quantity is preferable to an approach based on summing the
free energy changes associated with the addition of single molecules
to a cluster, on the grounds that we avoid the possible compounding
of statistical errors. The computational costs of our current study
of argon clusters have been higher than those of more traditional
methods such as grand canonical Monte Carlo \cite{merikanto2007},
for the same level of accuracy, largely because of our exploration
of different protocols and our use of an explicit helium thermostat,
but these can be reduced with further development. A particularly
powerful variant of the disassembly scheme is to separate a cluster
into two subclusters under similar mechanical guidance, in order to
relate the distribution of work performed to a free energy of `mitosis',
essentially a difference in excess free energies between the initial
cluster and the two final subclusters. Such comparisons would be unfeasible
to perform in Monte Carlo. The calculations are not onerous and an
evaluation of the excess free energy of clusters of up to 128 water
molecules is to be reported \cite{Lau15}. Furthermore, the explicit
thermostat can be replaced by an implicit scheme. With such tools,
and guided by the experience developed in the current investigation
of argon, we intend to carry out studies of clusters of water, acids
and organic molecules, species that are particularly relevant to the
process of aerosol nucleation in the atmosphere.
\begin{acknowledgments}
Hoi Yu Tang was funded by a PhD studentship provided by the UK Engineering
and Physical Sciences Research Council. We thank Gabriel Lau and George
Jackson for important comments.
\end{acknowledgments}
\appendix

\section{\label{sec:2ar_test}Argon dimer separation}

We test the feasibility of the approach using two protocols of controlled
dimer separation. First, the guide particles are made to drift apart
with the tether strengths held constant, and then we allow the tethers
to tighten over the course of the process. We determine the manner
of dimer separation that leads to an accurate estimate of the free
energy change.

We start by evaluating the free energy of a tethered dimer of argon-like
atoms analytically. Particles are distinguishable in molecular dynamics
simulations since they carry labels, so we take this into account
in the analysis. The initial Hamiltonian of the dimer system is
\begin{align}
H_{i}^{{\rm dimer}}=\, & \frac{\boldsymbol{p}_{1}^{2}}{2m}+\frac{\boldsymbol{p}_{2}^{2}}{2m}+\frac{1}{2}\kappa_{i}\left(\boldsymbol{x}_{1}-\mathbf{X}_{1}\right)^{2}\label{eq:2ar_hamiltonian_init.}\\
+\, & \frac{1}{2}\kappa_{i}\left(\boldsymbol{x}_{2}-\mathbf{X}_{2}\right)^{2}+\Phi\left(|\boldsymbol{x}_{1}-\boldsymbol{x}_{2}|\right),\nonumber
\end{align}
where $m$ is the argon mass, and $\Phi(\vert\boldsymbol{x}_{1}-\boldsymbol{x}_{2}\vert)$
is a pairwise interaction potential. When the guide particles both
lie at the origin $(\mathbf{X}_{1}=\mathbf{X}_{2}=0)$, the initial
partition function is
\begin{eqnarray}
 &  & Z_{i}^{{\rm dimer}}=\frac{1}{h^{6}}\int\exp\left(-\frac{p_{1}^{2}+p_{2}^{2}}{2mkT}\right)d\boldsymbol{p}_{1}d\boldsymbol{p}_{2}\label{eq:2ar_part.fn_init.}\\
 &  & \times\int\exp\left(-\kappa_{i}\frac{x_{1}^{2}+x_{2}^{2}}{2kT}\right)\exp\left(-\frac{\Phi\left(|\boldsymbol{x}_{1}-\boldsymbol{x}_{2}|\right)}{kT}\right)d\boldsymbol{x}_{1}d\boldsymbol{x}_{2},\nonumber
\end{eqnarray}
noting that there is no correction factor of one half since the atoms
are distinguishable. Substituting $\boldsymbol{r}=\boldsymbol{x}_{1}-\boldsymbol{x}_{2}$
and $\boldsymbol{R}=\boldsymbol{x}_{1}+\boldsymbol{x}_{2}$, the partition
function $Z_{i}^{{\rm dimer}}$ becomes
\begin{eqnarray}
 &  & \frac{1}{\lambda_{{\rm th}}^{6}}\int\frac{1}{8}\exp\left(-\kappa_{i}\frac{x_{1}^{2}+x_{2}^{2}}{2kT}\right)\exp\left(-\frac{\Phi\left(|\boldsymbol{x}_{1}-\boldsymbol{x}_{2}|\right)}{kT}\right)d\boldsymbol{r}d\boldsymbol{R}\nonumber \\
 &  & =\frac{1}{\lambda_{{\rm th}}^{6}}\frac{\pi}{2}\int\exp\left(-\kappa_{i}\frac{R^{2}+r^{2}}{4kT}\right)\exp\left(-\frac{\Phi\left(r\right)}{kT}\right)r^{2}drd\boldsymbol{R}\nonumber \\
 &  & =\frac{1}{\lambda_{{\rm th}}^{6}}\frac{\pi}{2}\left(\frac{4\pi kT}{\kappa_{i}}\right)^{\frac{3}{2}}\int_{0}^{r_{c}}r^{2}\exp\left(-\frac{\kappa_{i}r^{2}+4\Phi(r)}{4kT}\right)dr,\label{eq:2ar_part.fn_init.2}
\end{eqnarray}
where $\lambda_{{\rm th}}=h/(2\pi mkT)^{1/2}$ is the thermal de Broglie
wavelength. We have imposed an upper limit $r_{c}$ on the separation
between the two atoms, corresponding to a definition of what we mean
by a dimer.

For the final state in which the two argon atoms are tethered to respective
guide particles that are far apart, the Hamiltonian is simply that
in Eq. \eqref{eq:2ar_hamiltonian_init.} without the interaction term,
and with a final tether strength $\kappa_{f}$. The corresponding
final partition function is
\begin{eqnarray}
 &  & Z_{f}^{{\rm dimer}}=\frac{1}{h^{6}}\int\exp\left(-\frac{p_{1}^{2}+p_{2}^{2}}{2mkT}\right)d\boldsymbol{p}_{1}d\boldsymbol{p}_{2}\label{eq:2ar_part.fn_final}\\
 &  & \times\int\exp\left(-\kappa_{f}\frac{x_{1}^{2}+x_{2}^{2}}{2kT}\right)d\boldsymbol{x}_{1}d\boldsymbol{x}_{2}=\frac{1}{\lambda_{{\rm th}}^{6}}\left(\frac{2\pi kT}{\kappa_{f}}\right)^{3}.\nonumber
\end{eqnarray}
The free energy change in separating a dimer of tethered atoms can
therefore be expressed as
\begin{eqnarray}
 &  & \Delta F=kT\ln\left(Z_{i}^{{\rm dimer}}/Z_{f}^{{\rm dimer}}\right)\label{eq:2ar_DF.expression}\\
 &  & =kT\ln\left[\left(\frac{\kappa_{f}^{2}}{\kappa_{i}kT}\right)^{\frac{3}{2}}\int_{0}^{r_{c}}\frac{r^{2}}{2\sqrt{\pi}}\exp\left(-\frac{\kappa_{i}r^{2}+4\Phi(r)}{4kT}\right)dr\right],\nonumber
\end{eqnarray}
which can be evaluated numerically. The parameter $r_{c}$ is the
Stillinger radius used to identify a dimer configuration in the equilibrated
molecular dynamics simulation, to which we now turn.

We place two argon-like particles within a periodic cell with edge
length 50~Å, each tethered to guide particles through a harmonic
interaction $\frac{1}{2}\kappa(t)r^{2}$, where $r$ is the separation
between the argon atom and its guide, and $\kappa(t)$ is the tethering
force constant. The argon atoms are thermalised through interaction
with a gas of 100 helium-like atoms kept at constant temperature using
a Nosé-Hoover thermostat. Conventional masses of 39.85 and 4.003 amu
for the argon and helium-like particles are adopted, while the guide
particles are assigned a vastly greater mass of $4\times10^{12}$
amu. Interaction potentials are specified by
\begin{equation}
\Phi\left(r_{jk}\right)=4\epsilon_{jk}\left[\left(\frac{\sigma_{jk}}{r_{jk}}\right)^{12}-\left(\frac{\sigma_{jk}}{r_{jk}}\right)^{6}\right],\label{eq:LJ-pot-1}
\end{equation}
with parameters shown in Table \ref{tab:LJ-parameters1}, though it
should be noted that only the repulsive part of the interaction between
argon and helium is employed in order to prevent any binding between
the two. Simulations are performed at a temperature of 15~K such
that dimers are long-lived and a sufficient number of configurations
satisfying the separation criterion $r\le r_{c}=1.5\sigma_{{\rm ArAr}}$
can be obtained from the equilibrated trajectory. With a constant
tethering force constant of $0.05\,\text{kJ mol}^{-1}\text{Å}^{-2}$,
we generate an equilibrated molecular dynamics trajectory of duration
100~ns and choose $10^{3}$ dimer configurations for use as starting
points for the separation process.

\begin{table}
\centering{}%
\begin{tabular}{cccc}
\hline
$j$ & $k$ & $\epsilon_{jk}$ / $\mbox{kJ mol}^{-1}$ & $\sigma_{jk}$ / Å\tabularnewline
\hline
\hline
Ar & Ar & 0.995581 & 3.405\tabularnewline
He & He & 0.084311 & 2.600\tabularnewline
Ar & He & 0.289721 & 3.000\tabularnewline
\hline
\end{tabular}\protect\caption{\label{tab:LJ-parameters1}Parameters for the Lennard-Jones potentials,
where $j$ and $k$ are the atomic labels, $\epsilon_{jk}$ is the
depth of the potential well, and $\sigma_{jk}$ is the range parameter
\cite{hirshfelder64}.}
\end{table}

\subsubsection{Guiding at constant tether strength\label{sub:2Ar_LJ_tether.drag}}

\begin{figure}
\centering{}\includegraphics[width=1\columnwidth]{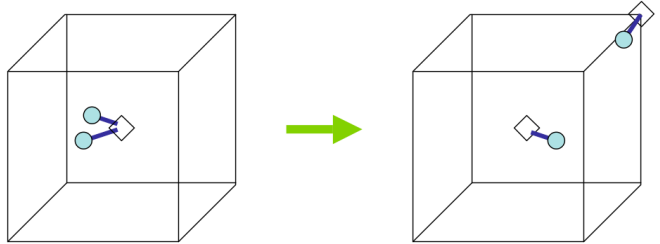}\protect\caption{\label{fig:2ar_sep_diagram} Illustration of the dimer separation
process. Both guide particles (diamonds) are initially at the origin,
but one is made to drift towards a corner of the simulation cell. }
\end{figure}

One of the guide particles drifts from the origin to a corner of the
cubic simulation cell over a separation time $t_{{\rm sep}}$ while
the other remains stationary (see Figure \ref{fig:2ar_sep_diagram}).
We choose $t_{{\rm sep}}$ to be 1, 2 or 4~ns and the velocity of
the moving guide particle (labelled 1) is given by $\mathbf{V}_{1}=[\mathbf{X}_{1}(t=t_{{\rm sep}})-\mathbf{X}_{1}(t=0)]/t_{{\rm sep}}$.

\begin{figure}
\begin{centering}
\includegraphics[width=1\columnwidth]{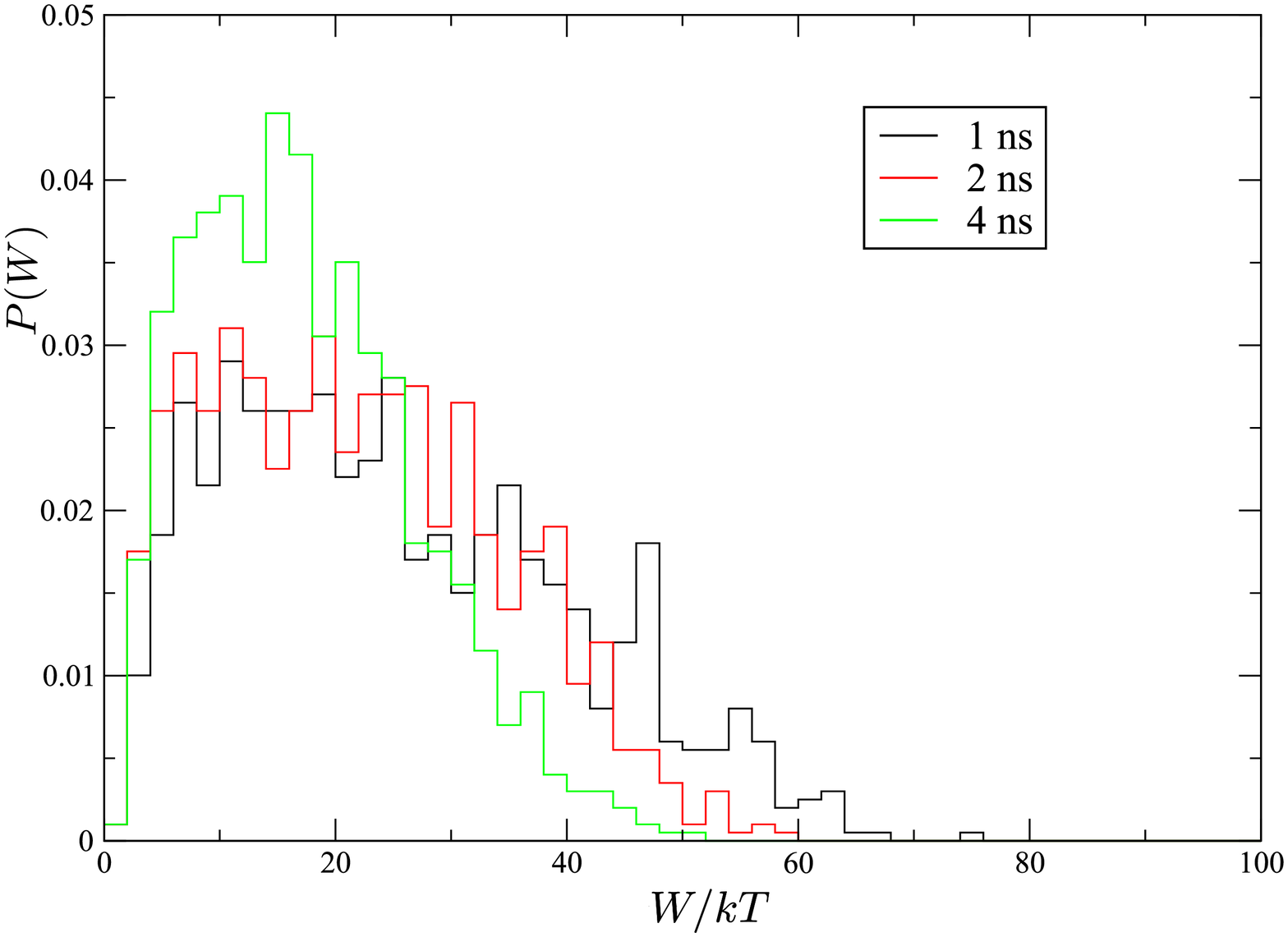}
\par\end{centering}

\protect\caption{\label{fig:2Ar_work_dist_rep.He_guidedragging} Distributions of the
work done in the disassembly of a dimer for separation times $t_{{\rm sep}}$
of 1, 2 and 4 ns. }
\end{figure}

For initial and final tethering force constants of $0.05\,\text{kJ mol}^{-1}\text{Å}^{-2}$,
the expected free energy change in separating the dimer is $5.716\, kT$
according to Eq. \eqref{eq:2ar_DF.expression}. Distributions of the
work done for each rate of dimer separation are shown in Figure \ref{fig:2Ar_work_dist_rep.He_guidedragging},
and the corresponding estimates of the free energy change obtained
from the Jarzynski equality are compared with the expected value in
the lower part of Figure \ref{fig:2Ar_DF,<W>_rep.He_guidedragging}.
A longer separation time leads to a better estimate of the free energy
change since the process is then closer to being quasistatic.

\begin{figure}
\begin{centering}
\includegraphics[width=1\columnwidth]{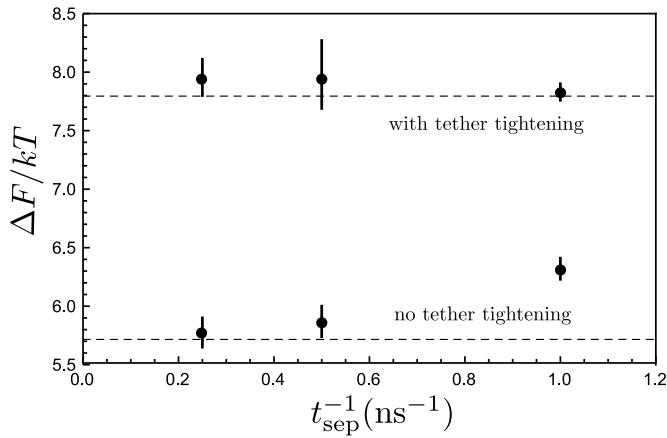}
\par\end{centering}

\protect\caption{\label{fig:2Ar_DF,<W>_rep.He_guidedragging}Convergence of the Jarzynski-estimated
free energy change toward the expected value (dashed line) as the
dimer separation rate is decreased, while keeping the tethering strength
constant (lower set) and when the tethers are tightened (upper set).}
\end{figure}

\subsubsection{Guiding with tether tightening}

\begin{figure}
\begin{centering}
\includegraphics[width=1\columnwidth]{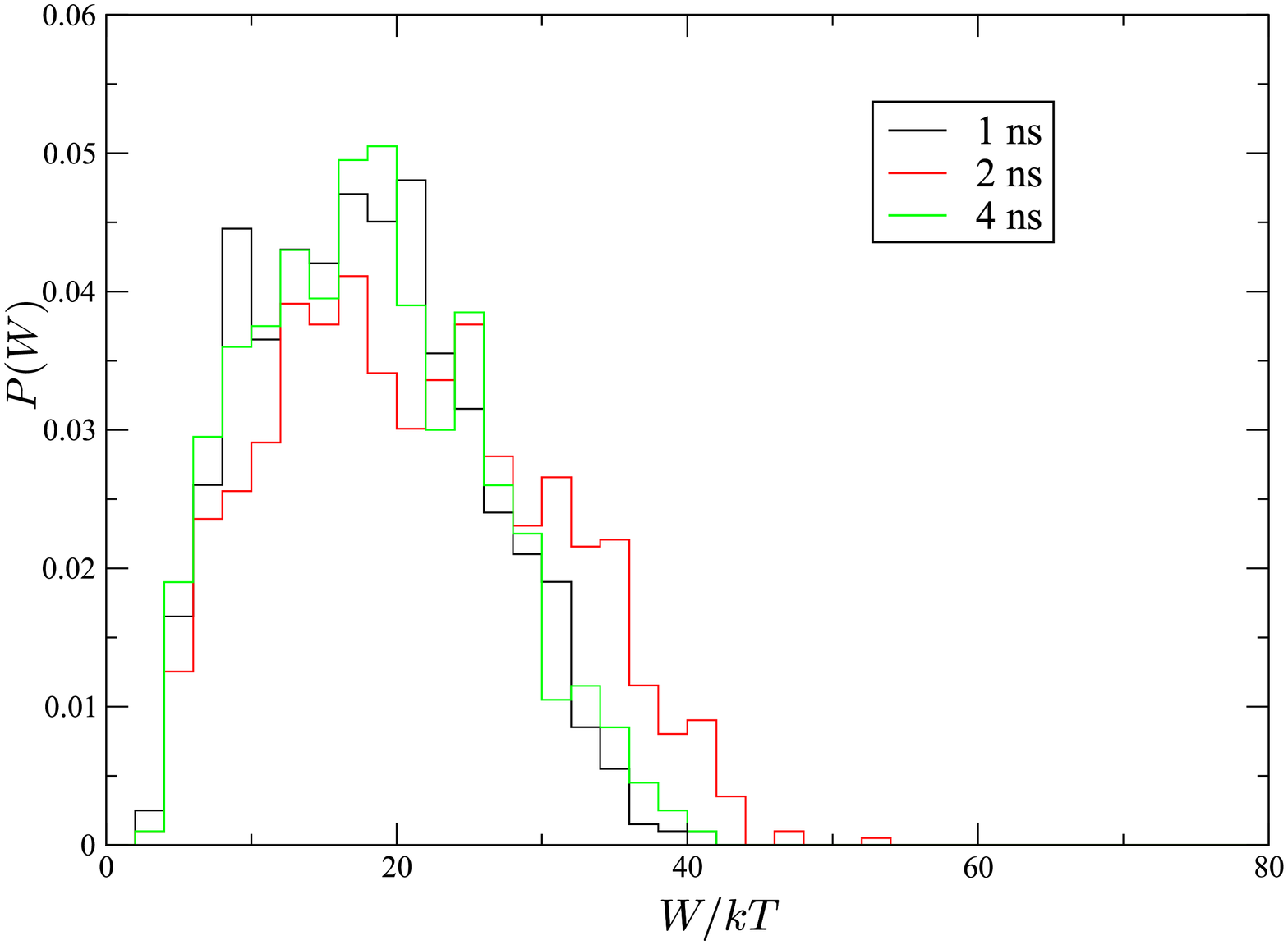}
\par\end{centering}

\protect\caption{\label{fig:2Ar_work_dist_rep.He_tdepk_Xdrag}Distributions of the
work of dimer disassembly where the atoms are guided apart and the
tethers tightened for three different separation times. }
\end{figure}

We now elaborate the process by tightening the tethers during guide
drift according to

\begin{eqnarray}
\kappa(t) & = & \kappa_{i}\quad\text{for}\; t\le t_{i}\nonumber \\
 & = & \kappa_{i}+\frac{\kappa_{f}-\kappa_{i}}{2}\left[1-\cos\left(\pi\frac{t-t_{i}}{t_{s}-t_{i}}\right)\right]\;\text{for}\; t_{i}<t\le t_{s}\nonumber \\
 & = & \kappa_{f}\quad\text{for}\; t>t_{s},\label{eq:spring_mod-1}
\end{eqnarray}
where $t_{i}$ is the time at which the force constant begins to change,
and $t_{s}$ is the time at which it reaches its final value. Once
again starting with dimer configurations and an initial tethering
force constant of $0.05\,\text{kJ mol}^{-1}\text{Å}^{-2}$ at 15~K,
three dimer separation times are investigated, during which the force
constant rises by a factor of two. The times $t_{i}$ and $t_{s}$
are specified as 20\% and 80\% of the total separation time. The expected
free energy change associated with dimer separation is $7.795\, kT$
according to Eq. \eqref{eq:2ar_DF.expression}. It can be seen from
the upper part of Figure \ref{fig:2Ar_DF,<W>_rep.He_guidedragging}
that all three separation rates give acceptable estimates of the free
energy change. Furthermore, the greater compatibility between the
distributions of the work performed at different separation rates
shown in Figure \ref{fig:2Ar_work_dist_rep.He_tdepk_Xdrag}, compared
with those in the simulations with constant tether strength, suggests
that a protocol where the tethers tighten while the guide particles
drift apart is more effective. Intuitively, the separation is then
conducted more firmly, and with less dissipation.

\section{Analysis of cluster free energies}

\subsection{Free and tethered clusters}

The canonical partition function $Z_{F}=\exp\left(-F_{F}/kT\right)$
for an untethered, or `free' cluster of $i$ indistinguishable particles
governed by a Hamiltonian $H$ composed of kinetic energy terms and
pairwise interactions is given by

\begin{eqnarray}
Z_{F} & = & \frac{1}{i!h^{3i}}\int\prod\limits _{j=1}^{i}d\boldsymbol{x}_{j}d\boldsymbol{p}_{j}\exp\left[-H\left(\left\{ \boldsymbol{x}_{k}\right\} \right)/kT\right],\label{eq:fren1-1}
\end{eqnarray}
 where $F_{F}$ is the associated free energy. For a cluster tethered
to the origin, the Hamiltonian will include an additional set of harmonic
potentials, such that the partition function is
\begin{eqnarray}
Z_{T} & = & \exp\left(-F_{T}/kT\right)=\frac{1}{i!h^{3i}}\int\prod\limits _{j=1}^{i}d\boldsymbol{x}_{j}d\boldsymbol{p}_{j}\nonumber \\
 & \times & \exp\left[-\left(H\left(\left\{ \boldsymbol{x}_{k}\right\} \right)+\sum\limits _{j=1}^{i}\frac{1}{2}\kappa_{i}x_{j}^{2}\right)/kT\right],\label{eq:fren2}
\end{eqnarray}
 where $F_{T}$ is the free energy of the tethered cluster, and $\kappa_{i}$
is the initial tethering force constant.

We insert a factor of unity in the form $1=\int\delta\left(\frac{1}{i}\sum_{j=1}^{i}\boldsymbol{x}_{j}-\boldsymbol{x}_{c}\right)d\boldsymbol{x}_{c}$
into Eqs. \eqref{eq:fren1-1} and \eqref{eq:fren2}, and transform
to particle coordinates with respect to the cluster centre of mass
$\boldsymbol{x}_{c}$, namely $\boldsymbol{x}_{j}^{\prime}=\boldsymbol{x}_{j}-\boldsymbol{x}_{c}$.
The partition function for a free cluster becomes
\begin{eqnarray}
Z_{F} & = & \frac{1}{i!h^{3i}}\int\prod\limits _{j=1}^{i}d\boldsymbol{x}_{j}^{\prime}d\boldsymbol{p}_{j}d\boldsymbol{x}_{c}\exp\left[-H\left(\left\{ \boldsymbol{x}_{k}^{\prime}\right\} \right)/kT\right]\nonumber \\
 &  & \times\delta\left(\frac{1}{i}\sum\limits _{j=1}^{i}\boldsymbol{x}_{j}^{\prime}\right)\nonumber \\
 & = & \frac{V}{i!h^{3i}}\int\prod\limits _{j=1}^{i}d\boldsymbol{x}_{j}^{\prime}d\boldsymbol{p}_{j}\exp\left[-H\left(\left\{ \boldsymbol{x}_{k}^{\prime}\right\} \right)/kT\right]\nonumber \\
 &  & \times\delta\left(\frac{1}{i}\sum\limits _{j=1}^{i}\boldsymbol{x}_{j}^{\prime}\right)=VZ_{F}^{c},\label{eq:fren4}
\end{eqnarray}
where $V$ is the system volume and $Z_{F}^{c}$ is the partition
function for a cluster whose centre of mass is fixed at the origin.
It should be noted that since the Hamiltonian contains pairwise interactions,
it may be rewritten as $H(\{\boldsymbol{x}_{k}\})=H(\{\boldsymbol{x}_{k}^{\prime}\})$
after the change of variables.

Similarly, the partition function for a tethered cluster can be rewritten
as
\begin{eqnarray}
 &  & Z_{T}=\frac{1}{i!h^{3i}}\int\prod\limits _{j=1}^{i}d\boldsymbol{x}_{j}^{\prime}d\boldsymbol{p}_{j}d\boldsymbol{x}_{c}\delta\left(\frac{1}{i}\sum\limits _{j=1}^{i}\boldsymbol{x}_{j}^{\prime}\right)\label{eq:fren5}\\
 &  & \times\exp\left[-\left(H\left(\left\{ \boldsymbol{x}_{k}^{\prime}\right\} \right)+\sum\limits _{j=1}^{i}\frac{1}{2}\kappa_{i}x_{j}^{2}\right)/kT\right].\nonumber
\end{eqnarray}
The second term in the exponent of Eq. \eqref{eq:fren5} may be simplified
using the constraint $\sum_{j=1}^{i}\boldsymbol{x}_{j}^{\prime}=0$
and it follows that $\sum_{j=1}^{i}x_{j}^{2}=\sum_{j=1}^{i}x_{j}^{\prime2}+ix_{c}^{2}$,
giving
\begin{eqnarray}
 &  & Z_{T}=\frac{1}{i!h^{3i}}\int\prod\limits _{j=1}^{i}d\boldsymbol{x}_{j}^{\prime}d\boldsymbol{p}_{j}d\boldsymbol{x}_{c}\exp\left[-\frac{1}{2}\kappa_{i}ix_{c}^{2}/kT\right]\nonumber \\
 &  & \times\exp\left[-\left(H\left(\left\{ \boldsymbol{x}_{k}^{\prime}\right\} \right)+\frac{1}{2}\kappa_{i}\sum\limits _{j=1}^{i}x_{j}^{\prime2}\right)/kT\right]\delta\left(\frac{1}{i}\sum\limits _{j=1}^{i}\boldsymbol{x}_{j}^{\prime}\right)\nonumber \\
 &  & =\left(\frac{2\pi kT}{i\kappa_{i}}\right)^{\frac{3}{2}}\frac{1}{i!h^{3i}}\int\prod\limits _{j=1}^{i}d\boldsymbol{x}_{j}^{\prime}d\boldsymbol{p}_{j}\exp\left[-H\left(\left\{ \boldsymbol{x}_{k}^{\prime}\right\} \right)/kT\right]\nonumber \\
 &  & \times\exp\left[-\frac{1}{2}\kappa_{i}\sum\limits _{j=1}^{i}x_{j}^{\prime2}/kT\right]\delta\left(\frac{1}{i}\sum\limits _{j=1}^{i}\boldsymbol{x}_{j}^{\prime}\right)\nonumber \\
 &  & =\left(\frac{2\pi kT}{i\kappa_{i}}\right)^{\frac{3}{2}}Z_{T}^{c},\label{eq:fren6}
\end{eqnarray}
where $Z_{T}^{c}$ is the partition function of a cluster constrained
to have its centre of mass at the origin as well as having its constituent
particles tethered to the origin by a harmonic potential.

Next, we employ the Gibbs-Bogoliubov approach \cite{hansen1986,ishihara1968}
to compare the free energies $F_{F}^{c}$ and $F_{T}^{c}$ of systems
with Hamiltonians $H_{0}$ and Hamiltonian $H_{0}+U$, defined by
$\exp\left(-F_{F}^{c}/kT\right)=\int d\Gamma\exp\left[-H_{0}/kT\right]$
and $\exp\left(-F_{T}^{c}/kT\right)=\int d\Gamma\exp\left[-\left(H_{0}+U\right)/kT\right]$,
where $\Gamma$ represents the configuration of a system, and $d\Gamma$
is proportional to the phase space volume element $\Pi_{j}d\boldsymbol{x}_{j}^{\prime}d\boldsymbol{p}_{j}$.
In the context of the tethered cluster described by Eq. \eqref{eq:fren6},
$U$ represents the term $\frac{1}{2}\kappa_{i}\sum_{j=1}^{i}x_{j}^{\prime2}$,
while $H_{0}$ is the untethered Hamiltonian $H\left(\left\{ \boldsymbol{x}_{k}^{\prime}\right\} \right)$
modified by the delta function constraint. $F_{T}^{c}$ is therefore
the free energy of a tethered cluster with its centre of mass further
constrained to lie at the origin, and is equal to $-kT\ln Z_{T}^{c}$.
A similar relationship exists between $F_{F}^{c}$, the free energy
of an untethered cluster with fixed centre of mass, and $Z_{F}^{c}$.

The free energies $F_{F}^{c}$ and $F_{T}^{c}$ may be related through
\begin{eqnarray}
\!\!\!\exp\left(-F_{T}^{c}/kT\right) & = & \frac{\int d\Gamma\exp\left(-H_{0}/kT\right)\exp\left(-U/kT\right)}{\int d\Gamma\exp\left(-H_{0}/kT\right)}\nonumber \\
 &  & \times\int d\Gamma\exp\left(-H_{0}/kT\right)\nonumber \\
 & = & \left<\exp\left(-U/kT\right)\right>_{0}\exp\left(-F_{F}^{c}/kT\right),\label{eq:fren9}
\end{eqnarray}
where angle brackets represent an average in the statistical ensemble
corresponding to $H_{0}$. For small $\left<U/kT\right>_{0}$, we
can write $\left<\exp\left(-U/kT\right)\right>_{0}\simeq\exp\left(-\left<U\right>_{0}/kT\right)$,
and hence
\begin{eqnarray}
\exp\left(-F_{T}^{c}/kT\right) & \simeq & \exp\left[\left(-F_{F}^{c}-\left<U\right>_{0}\right)/kT\right],\label{eq:fren11}
\end{eqnarray}
 with $\langle U\rangle_{0}$ given by
\begin{equation}
\langle U\rangle_{0}=\frac{\int d\Gamma U\left(\left\{ \boldsymbol{x}_{k}^{\prime}\right\} \right)\exp\left(-H_{0}/kT\right)}{\int d\Gamma\exp\left(-H_{0}/kT\right)}.\label{eq:fren11.1}
\end{equation}

$U(\{\boldsymbol{x}_{k}^{\prime}\})$ is a sum of single-particle
harmonic potentials of the form $U_{{\rm HO}}(\boldsymbol{x}_{k}^{\prime})=\frac{1}{2}\kappa_{i}x_{k}^{\prime2}$,
so Eq. \eqref{eq:fren11.1} can be written as
\begin{eqnarray}
\!\!\!\!\!\!\langle U\rangle_{0} & = & \frac{\sum_{k=1}^{i}\int d\Gamma U_{{\rm HO}}\left(\boldsymbol{x}_{k}^{\prime}\right)\exp\left(-H_{0}/kT\right)}{\int d\Gamma\exp\left(-H_{0}/kT\right)}\nonumber \\
 & = & i\frac{\int d\Gamma U_{{\rm HO}}\left(\boldsymbol{x}_{k}^{\prime}\right)\exp\left(-H_{0}/kT\right)}{\int d\Gamma\exp\left(-H_{0}/kT\right)}=i\langle U_{{\rm HO}}\rangle_{0}.\label{eq:fren12}
\end{eqnarray}
We next introduce the spatial density profile of a single particle
(labelled $k$ without loss of generality) in a cluster constrained
to have its centre of mass at the origin but not tethered, namely
\begin{equation}
\rho_{0}(\boldsymbol{y})=\frac{\int d\Gamma\exp\left(-H_{0}/kT\right)\delta(\boldsymbol{x}_{k}^{\prime}-\boldsymbol{y})}{\int d\Gamma\exp\left(-H_{0}/kT\right)},\label{eq:fren13}
\end{equation}
 with $\int\rho_{0}(\boldsymbol{y})d\boldsymbol{y}=1$. We can write
\begin{equation}
\left<U_{{\rm HO}}\right>_{0}=\int\rho_{0}(\boldsymbol{y})U_{{\rm HO}}(\boldsymbol{y})d\boldsymbol{y},\label{eq:fren14}
\end{equation}
which represents the average tethering energy of a particle that is
spatially distributed according to the density $\rho_{0}(\boldsymbol{y})$.
The condition that the tether potential makes a relatively small contribution
to the mean energy of the cluster is $\langle U_{{\rm HO}}\rangle_{0}=\frac{1}{2}\kappa_{i}\int\rho_{0}(\boldsymbol{y})y^{2}d\boldsymbol{y}\ll kT$,
in which case the approximations involved in the Gibbs-Bogoliubov
approach are acceptable and the initial tethering potential weak enough
that the cluster is only slightly distorted in comparison with a free
cluster. Thus we write
\begin{equation}
Z_{T}^{c}=\exp\left(-F_{T}^{c}/kT\right)\simeq\exp\left[\left(-F_{F}^{c}-i\left<U_{{\rm HO}}\right>_{0}\right)/kT\right].\label{eq:fren15}
\end{equation}
Eq. \eqref{eq:fren6} can then be written as
\begin{eqnarray}
 &  & Z_{T}=\left(\frac{2\pi kT}{i\kappa_{i}}\right)^{\frac{3}{2}}\frac{1}{i!h^{3i}}\int\prod_{j=1}^{i}d\boldsymbol{x}_{j}^{\prime}d\boldsymbol{p}_{j}\exp\left[-H\left(\left\{ \boldsymbol{x}_{k}^{\prime}\right\} \right)/kT\right]\nonumber \\
 &  & \times\exp\left[-i\int\rho_{0}\left(\boldsymbol{y}\right)\kappa_{i}y^{2}d\boldsymbol{y}/2kT\right]\delta\left(\frac{1}{i}\sum_{j=1}^{i}\boldsymbol{x}_{j}^{\prime}\right),\label{eq:fren16}
\end{eqnarray}
such that the relationship between the partition function of a tethered
cluster, and the partition function of a free cluster with a constrained
centre of mass $Z_{F}^{c},$ is
\begin{equation}
Z_{T}=Z_{F}^{c}\left(\frac{2\pi kT}{i\kappa_{i}}\right)^{\frac{3}{2}}\exp\left[-i\int\rho_{0}\left(\boldsymbol{y}\right)\kappa_{i}y^{2}d\boldsymbol{y}/2kT\right].\label{eq:fren17}
\end{equation}
Combining Eqs. \eqref{eq:fren4} and \eqref{eq:fren17} then gives
\begin{eqnarray}
\!\!\!\!\!\!\!\!\!\!\!\!\!\ln Z_{T} & = & \ln\!\left[\frac{Z_{F}}{V}\left(\frac{2\pi kT}{i\kappa_{i}}\right)^{\frac{3}{2}}\right]-\frac{i\kappa_{i}}{2kT}\int\rho_{0}(\boldsymbol{y})y^{2}d\boldsymbol{y},\label{eq:fren18}
\end{eqnarray}
or
\begin{equation}
F_{F}-F_{T}=-kT\ln\left[\rho_{c}\left(0\right)V\right]-\frac{i\kappa_{i}}{2}\int\rho_{0}(\boldsymbol{y})y^{2}d\boldsymbol{y},\label{eq:fren19}
\end{equation}
where $\left(i\kappa_{i}/2\pi kT\right)^{3/2}$ has been replaced
by a function $\rho_{c}(0)$, representing the probability density
that the centre of mass of the tethered cluster lies at the origin.
This equivalence can be demonstrated by deriving the distribution
of the cluster centre of mass, through considering a single particle
with mass $M=im$ and coordinates $\boldsymbol{x}_{c}$ and $\boldsymbol{p}_{c}$
residing in a potential $i\kappa_{i}x_{c}^{2}/2$. The positional
probability density at $\boldsymbol{z}$ is
\begin{eqnarray}
\rho_{c}(\boldsymbol{z}) & = & \frac{\int d\boldsymbol{x}_{c}d\boldsymbol{p}_{c}\exp\left(-\frac{i\kappa_{i}x_{c}^{2}}{2kT}-\frac{p_{c}^{2}}{2MkT}\right)\delta\left(\boldsymbol{x}_{c}-\boldsymbol{z}\right)}{\int d\boldsymbol{x}_{c}d\boldsymbol{p}_{c}\exp\left(-\frac{i\kappa_{i}x_{c}^{2}}{2kT}-\frac{p_{c}^{2}}{2MkT}\right)}\nonumber \\
 & = & \left(\frac{i\kappa_{i}}{2\pi kT}\right)^{3/2}\exp\left(-\frac{i\kappa_{i}z^{2}}{2kT}\right),\label{eq:fren20}
\end{eqnarray}
 such that $\rho_{c}(0)=\left(i\kappa_{i}/(2\pi kT)\right)^{3/2}$.

The purpose of the substitution is that the first term on the right
hand side in Eq. \eqref{eq:fren19} may be interpreted as two competing
contributions to the free energy difference $F_{F}-F_{T}$. We write
\begin{equation}
-kT\ln\left[\rho_{c}\left(0\right)V\right]=-T\left[-k\ln\left(\frac{1}{\rho_{c}(0)}\right)+k\ln V\right],\label{eq:fren19.1}
\end{equation}
 such that the first term corresponds to the removal of the entropic
contribution to free energy associated with the freedom of motion
of the cluster centre of mass within a constrained volume $1/\rho_{c}(0)$,
brought about by the tethers, and the second term represents the addition
of entropic free energy corresponding to the freedom of motion in
volume $V$. Finally, the second term in Eq. \eqref{eq:fren19} is
an estimate of the removal of tethering potential energy when relating
a tethered to a free cluster.

\subsection{Excess free energy from the free energy of disassembly}

We now establish the relationship between the free energy of a free
cluster to the cluster work of formation defined as $\phi\left(i\right)=\Omega_{s}(i)-ikT\ln S$,
where $\Omega_{s}(i)=F_{F}(i)-i\mu_{s}$ is the grand potential of
a free cluster of $i$ particles in an environment at chemical potential
$\mu_{s}$ for which the bulk condensed and vapour phases coexist.
The excess free energy (difference) of the cluster is therefore
\begin{align}
F_{s}\left(i\right) & =\phi\left(i\right)-\phi(1)+\left(i-1\right)kT\ln S\nonumber \\
 & =F_{F}\left(i\right)-F\left(1\right)-\left(i-1\right)\mu_{s},\label{eq:21b}
\end{align}
having used Eq. \eqref{eq:nucleation_barrier_contributions}.

Assuming the vapour is ideal, the coexistence chemical potential $\mu_{s}$
and the monomer Helmholtz free energy $F(1)$ are simply $\mu_{s}=kT\ln(\rho_{vs}\Lambda)$
and $F(1)=-kT\ln(V/\Lambda)$, respectively, where $\rho_{vs}$ is
the particle density in a saturated vapour and $\Lambda=\lambda_{{\rm th}}^{3}$
with $\lambda_{{\rm th}}=h/(2\pi mkT)^{1/2}$. The excess free energy
$F_{s}(i)$ can now be expressed as
\begin{eqnarray}
F_{s}(i) & = & F_{F}+kT\ln\left(V/\Lambda\right)-\left(i-1\right)kT\ln\left(\rho_{vs}\Lambda\right)\nonumber \\
 & = & F_{T}-kT\ln\left[\rho_{c}\left(0\right)V\right]-\frac{i\kappa_{i}}{2}\int\rho_{0}\left(\boldsymbol{y}\right)y^{2}d\boldsymbol{y}\nonumber \\
 &  & +kT\ln\left(V/\Lambda\right)-\left(i-1\right)kT\ln\left(\rho_{vs}\Lambda\right).\label{eq:fren23}
\end{eqnarray}

Now we consider the free energy change associated with the process
of cluster disassembly. The difference in free energy between separated
constituent particles each tethered to a guide particle, and a tethered
cluster, is $\delta F=F_{f}-F_{T}$, where $F_{f}=-3ikT\ln\left(kT/\hbar\omega_{f}\right)$
is the free energy of $i$ harmonic oscillators in three dimensions,
where the angular frequency $\omega_{f}=\left(\kappa_{f}/m\right)^{1/2}$
of the oscillators is related to the final value of the tethering
force constant $\kappa_{f}$ .

It should be recognised, however, that the quantity $\delta F$ is
\emph{not} the free energy difference extracted from the molecular
dynamics simulations of cluster disassembly. Molecular dynamics simulations
always involve distinguishable particles, since they are assigned
labels, and $\delta F$ is a difference between the free energy of
$i$ indistinguishable particles in a cluster, and $i$ particles
that are distinguishable through having been physically separated
to regions around their final tether points.

The free energy difference that is extracted in our procedure is actually
$\Delta F=F_{f}-F_{T}^{{\rm dist}}$, where the superscript in $F_{T}^{{\rm dist}}$
reminds us that it is the free energy of a tethered cluster of distinguishable
particles. But we can relate the partition function of such a cluster
to the partition function $Z_{T}$ for indistinguishable particles
by the usual classical procedure, namely $Z_{T}^{{\rm dist}}=i!Z_{T}$,
and since $F_{T}^{{\rm dist}}=-kT\ln Z_{T}^{{\rm dist}}=-kT\ln Z_{T}-kT\ln i!=F_{T}-kT\ln i!$
we have
\begin{equation}
\Delta F=F_{f}-F_{T}+kT\ln i!=\delta F+kT\ln i!,\label{eq:24b}
\end{equation}
such that $F_{T}=F_{f}-\delta F=F_{f}-\Delta F+kT\ln i!$. Substituting
into Eq. \eqref{eq:fren23} then gives
\begin{align}
F_{s}(i) & =-\Delta F-ikT\ln\left(\rho_{vs}v_{{\rm HO}}\right)+kT\ln i!\nonumber \\
 & -kT\ln\left(\frac{\rho_{c}(0)}{\rho_{vs}}\right)-\frac{i\kappa_{i}}{2}\int\rho_{0}\left(\boldsymbol{y}\right)y^{2}d\boldsymbol{y},\label{eq:fren25}
\end{align}
where $v_{{\rm HO}}=\left(2\pi kT/\kappa_{f}\right)^{3/2}$ is a volume
scale associated with the confinement of particles within the final
harmonic tether potentials. It should be noted that the excess free
energy $F_{s}$ does not depend upon the Planck constant $h$, nor
on the system volume $V$, as is to be expected.

In order to complete our specification of $F_{s}(i)$ in terms of
$\Delta F$ and material properties, we need to estimate the final
term in Eq. \eqref{eq:fren25}. We write $\int\rho_{0}(\boldsymbol{y})y^{2}d\boldsymbol{y}=\int_{0}^{\infty}\rho_{0}(r)4\pi r^{4}dr$,
where $r$ is the distance from the cluster centre of mass, and recall
that $\rho_{0}(r)$ is the single-particle density profile in an untethered
cluster with fixed centre of mass. As an approximation, we imagine
the cluster to be spherical with a constant particle density, such
that $\rho_{0}(r)\simeq\rho_{l}/i$ for $0<r<r_{{\rm max}}$, where
$\rho_{l}$ is the particle density in the condensed phase, and $r_{{\rm max}}$
is the radius of the cluster. Since the probability density $\rho_{0}(r)$
is normalised, we have $\int_{0}^{r_{{\rm max}}}(\rho_{l}/i)4\pi r^{2}dr=1$,
such that $r_{{\rm max}}=\left(3i/4\pi\rho_{l}\right)^{1/3}$ and
so
\begin{eqnarray}
\int\limits _{0}^{r_{{\rm max}}}\frac{\rho_{l}}{i}4\pi r^{4}dr & = & \frac{4\pi\rho_{l}}{5i}r_{{\rm max}}^{5}=\frac{3}{5}\left(\frac{3iv_{l}}{4\pi}\right)^{2/3},\label{eq:fren29}
\end{eqnarray}
where $v_{l}=1/\rho_{l}$ is the volume per particle in the condensed
phase. Substituting this into Eq. (\ref{eq:fren25}) gives Eqs. (\ref{eq:fren30_breakdown1}-\ref{eq:fren30_breakdown5})
in the main text.

%
\end{document}